\documentclass[%
 reprint,
superscriptaddress,
 amsmath,amssymb,
 aps
]{revtex4-2}

\usepackage{graphicx                                }
\usepackage{dcolumn}
\usepackage{bm}
\usepackage{hyperref}
\hypersetup{
    colorlinks,
    linkcolor={red},
    citecolor={blue},
    urlcolor={blue}
}

\newcommand{\iu}{{i\mkern1mu}}

\DeclareMathOperator{\sign}{sign}

\begin{document}

\title{Two identical 1D anyons with zero-range interactions: Exchange statistics,
scattering theory, and anyon-anyon mapping}
\author{Ra\'ul Hidalgo-Sacoto}
\affiliation{%
 Quantum Systems Unit, Okinawa Institute of Science and Technology Graduate University, \\ Onna, Okinawa 904-0495, Japan}%
\author{Thomas Busch}
\affiliation{%
 Quantum Systems Unit, Okinawa Institute of Science and Technology Graduate University, \\ Onna, Okinawa 904-0495, Japan}%
\author{D. Blume}
\affiliation{Homer L. Dodge Department of Physics and Astronomy,
  The University of Oklahoma,
  440 W. Brooks Street,
  Norman,
Oklahoma 73019, USA}
\affiliation{Center for Quantum Research and Technology,
  The University of Oklahoma,
  440 W. Brooks Street,
  Norman,
Oklahoma 73019, USA}
\date{\today}

\begin{abstract}
While elementary particles obey either bosonic or fermionic exchange statistics,
generalized exchange statistics that interpolate between bosons and fermions---applicable to quasi-particles--- constitute an intriguing topic,
both from the fundamental and practical points of view.
This work develops a scattering framework for two identical 1D bosonic anyons and 
two identical 1D fermionic anyons
with zero-range contact interactions.
The two-body system with zero-range interactions, both in free space and under external confinement, is used to illustrate the recently proposed bosonic-anyon---fermionic-anyon mapping~(R. Hidalgo-Sacoto {\em{et al.}},  arXiv:2505.17669),
which connects the eigenstates of bosonic anyons to those of fermionic anyons and vice versa. 
Performing explicit calculations for two-particle systems, the
momentum distributions and the off-diagonal correlations of the single-particle density matrix for bosonic anyons and fermionic anyons are 
confirmed to be distinct. 
We also 
confirm the previously derived asymptotic coefficients of the momentum distribution tail at orders $k^{-2}$ and $k^{-3}$
for
two harmonically confined anyons.
 Non-universal contributions at order $k^{-4}$ are discussed.
\end{abstract}
\maketitle

\section{Introduction}

Quantum statistics, i.e., the behavior upon exchange of identical quantum particles, has a profound impact on the
system behavior in the regime where the de Broglie wavelength is comparable to or larger than the
interparticle spacing~\cite{leinaas1977theory}.
Prominent examples are the Bose-Einstein and Fermi-Dirac distribution functions, which govern
the behaviors of a collection of identical bosons and identical fermions, respectively~\cite{baym2018lectures}.
Conventionally, particles in 1D are categorized as either
being a (composite) boson or a (composite) fermion~\cite{giamarchi2003quantum}. 
There do, however, exist examples of exchange statistics that do not fall into the boson- or fermion-categories, 
fractional statistics in 2D being one such example~\cite{halperin1984statistics, arovas1984fractional, wilczek1990fractional, haldane1991fractional, stern2008anyons}. 
In addition to the exchange statistics, particle interactions can have a tremendous effect on
the system properties~\cite{lieb1963exact, yang1967some, bloch2008many}. In 1D, the Bose-Fermi mapping transforms---provided the bosonic and fermionic scattering lengths are equal to each other---the wavefunction of $N$ identical bosons with two-body even-parity zero-range interactions characterized by the 1D scattering length $a_+$ into the wavefunction of $N$ identical fermions with two-body odd-parity zero-range interactions characterized by the 1D scattering length $a_-$~\cite{cheon1998realizing, cheon1999fermion, girardeau2003fermi, girardeau2004effective}. This mapping is special to 1D 
systems that consist of $N$ identical particles with two-body zero-range interactions~\cite{lieb1963exact}. 

Over the past decade, renewed interest in non-standard exchange statistics has developed~\cite{stern2008anyons, kwan2024realization, wang2025particle}. The resurgence is, at least 
partially, driven by the unprecedented control that is afforded by atomic systems such as
ultracold atoms loaded into an optical lattice~\cite{lewenstein2012ultracold} or atomic tweezer arrays~\cite{kaufman2021quantum} and ions held in Paul traps~\cite{tomza2019cold}. 
Examples of unconventional exchange statistics are para-statistics~\cite{green1953generalized}, which can be formulated in any spatial dimension
 in
terms of higher-dimensional
representations of the permutation group~\cite{wang2025particle}, 
and anyonic statistics~\cite{wilczek1990fractional, haldane1991fractional}, which was originally formulated for two-dimensional systems in terms of
the braid group~\cite{nakamura2020direct}. 

This work considers anyon-like exchange statistics of 1D systems in the continuum ~\cite{kundu1999exact, girardeau2006anyon, batchelor2006one},
in which the wave function acquires a phase upon exchange of two identical particles that depends on the statistical parameter $\alpha$, and the position coordinates of these particles. The statistical parameter interpolates between bosonic and fermionic exchange statistics.
While there exists a comparatively vast literature on studies of anyons in lattices~\cite{keilmann2011statistically, greschner2015anyon, munoz2020anyonic, bonkhoff2021bosonic, leonard2023realization,dhar2024anyonization, kwan2024realization}, the study of interacting anyons in the continuum is still in its infancy~\cite{kundu1999exact, girardeau2006anyon, batchelor2006one,del2008fermionization, zinner2015strongly, wang2024boson}.  
We investigate the simplest interacting continuum system~\cite{busch1998two}, namely two identical anyons in 1D, within the framework of non-relativistic quantum mechanics. 
Extending the standard 1D scattering framework~\cite{olshanii1998atomic, kanjilal2004nondivergent, kanjilal2009pseudopotential},
we define the anyonic scattering phase shift, and subsequently the anyonic scattering length, by introducing two 
linearly independent reference functions that are solutions to the free-particle 
Schr\"odinger equation and 
possess the desired generalized exchange statistics.
Using the anyonic scattering length, a zero-range two-body interaction potential
that supports anyonic solutions is identified. Our approach highlights the demands on the interaction potential 
for the two-particle solutions to be consistent with the anyonic exchange statistics in 1D. Our two-particle treatment provides a concrete example of the recently proposed bosonic anyon-fermionic anyon mapping~\cite{paper1}.

The remainder of this article is organized as follows. 
Section~\ref{sec_two_body_scattering} defines and analyzes the two-body scattering solutions for two identical anyons that interact through a zero-range pseudopotential. The constraints imposed by the anyonic particle exchange symmetry on the pseudopotential are discussed. 
Section~\ref{sec_fewbd_observables} defines relevant observables such as the momentum distribution, summarizes key aspects of the tail of the momentum distribution from our very recent publication~\cite{paper1}, and derives a number of symmetry properties that are inherently linked to the chirality of the 1D anyons under study.
Section~\ref{sec_examples} provides explicit results for two free anyons and two anyons under external confinement. Universal and non-universal properties are analyzed.
Finally,  Sec.~\ref{sec_conclusion} summarizes our results.

\section{Two-body scattering solutions}\label{sec_two_body_scattering}

\subsection{Basics}
\label{sec_intro_QDT_framework}
We consider two identical particles of mass $m$ with coordinates $z_1$ and $z_2$. The two-body interaction potential $V(z)$ is assumed to be a short-range interaction potential that depends only on the relative coordinate $z$, $z=z_1-z_2$, and not on the center-of-mass coordinate $Z$, $Z=(z_1+z_2)/2$, or spin degrees of freedom. 
It is assumed that $V(z)$ falls off faster than $|z|^{-1}$ as $|z| \rightarrow \infty$. Since the center-of-mass motion separates, 
we are interested in finding the scattering solutions for the time-independent relative Schr\"odinger
equation for a pseudo-particle with reduced mass $\mu$ ($\mu=m/2$),
\begin{eqnarray}
\label{eq_se_rel}
\left( -\frac{\hbar^2}{2 \mu} \frac{\partial^2}{\partial z^2} +V(z)
\right) \psi(z) = E \psi(z),
\end{eqnarray}
where $E$ denotes the relative energy.
Since $E \ge 0$ for the scattering solutions, it is
convenient to introduce the real wave vector $k$,
\begin{eqnarray}
\label{eq_wavevector}
k = \sqrt{\frac{2 \mu E}{\hbar^2} }.
\end{eqnarray}
The Schr\"odinger equation, Eq.~(\ref{eq_se_rel}), is a second-order differential equation in $z$. Correspondingly,
it has two linearly independent free-particle solutions
for each $k$, namely $\sin(kz)$ and $\cos(kz)$; normalization constants will be introduced later as needed.
Since any linear combination of these solutions is also a solution, the linearly independent solutions can be 
written in other ways [e.g., $\exp(\iu k z)$ and $\exp(-\iu k z)$ are also used frequently].
In scattering theory for systems with short-range interactions, the outside
solution, i.e., the solution for $|z| \ge z_0$ ($z_0$ is positive), is 
given in terms of a linear combination of the regular function $f_k(z)$
and the irregular  function $g_k(z)$, where the energy-dependent scattering phase shift $\delta_\mathrm{sc}(k)$ quantifies the admixture of the 
irregular function $g_k(z)$ due to the interaction potential~\cite{kanjilal2004nondivergent, kanjilal2009pseudopotential},
\begin{eqnarray}
\label{eq_outside}
\psi_k(z) \underset{|z| \ge z_0}{\longrightarrow}
\beta(k) \left[ f_k(z) + \tan(\delta_\mathrm{sc}(k)) g_k(z) \right];
\end{eqnarray}
here, $\beta(k)$ is an energy-dependent normalization coefficient. The linearly independent reference functions $f_k(z)$ and $g_k(z)$ are solutions to the time-independent free-particle Schr\"odinger equation, i.e., to the Schr\"odinger equation for 
vanishing interaction potential [$V(z)=0$]; as elaborated on below, their choice
depends on the particle exchange statistics. The 1D scattering length $a_{\text{sc}}$, 
which is 
independent of 
the 
wave vector $k$, is defined through 
\begin{eqnarray}
\label{eq_ascatt}
    a_{\mathrm{sc}} = 
     \lim_{k\rightarrow 0}
    -\frac{\tan(\delta_\mathrm{sc}(k))}{k}.
\end{eqnarray}

\subsection{Anyonic exchange statistics}
\label{sec_anyonic_exchange}
If $\chi(z)$ is an un-symmetrized relative wave function,
then a symmetric wave function $\psi_+(z)$ and an anti-symmetric wave function $\psi_-(z)$ can be constructed 
by applying the two-particle symmetrization and anti-symmetrization operators, respectively,
\begin{eqnarray}
\psi_{\pm}(z)= \frac{1}{\sqrt{2}} \left( 1 \pm \hat{P}_{12} \right) \chi(z);
\end{eqnarray}
here, $\hat{P}_{12}$ exchanges the coordinates $z_1$ and $z_2$ ($z$ goes to $-z$ under the $\hat{P}_{12}$ operation).
Acting with $\hat{P}_{12}$ on $\psi_{\pm}(z)$, one finds
\begin{eqnarray}
\label{eq_exchange_simple}
 \psi_{\pm}(-z)= \pm \psi_{\pm}(z) ,
\end{eqnarray}
where the multiplicative $\pm1$ factor on the right-hand side can be interpreted as a phase factor
[$+1=\exp(0)$ and $-1=\exp(-\iu\pi)$].
It is clear from Eq.~(\ref{eq_exchange_simple}) that $\psi_+(z)$ obeys bosonic exchange statistics while
$\psi_-(z)$ obeys fermionic exchange statistics. Note that we are justified to discuss the exchange  statistics by considering the relative coordinate only since the center-of-mass coordinate $Z$ is unchanged under application of $\hat{P}_{12}$.

\begin{figure}
    \centering
    \includegraphics[width=1\linewidth]{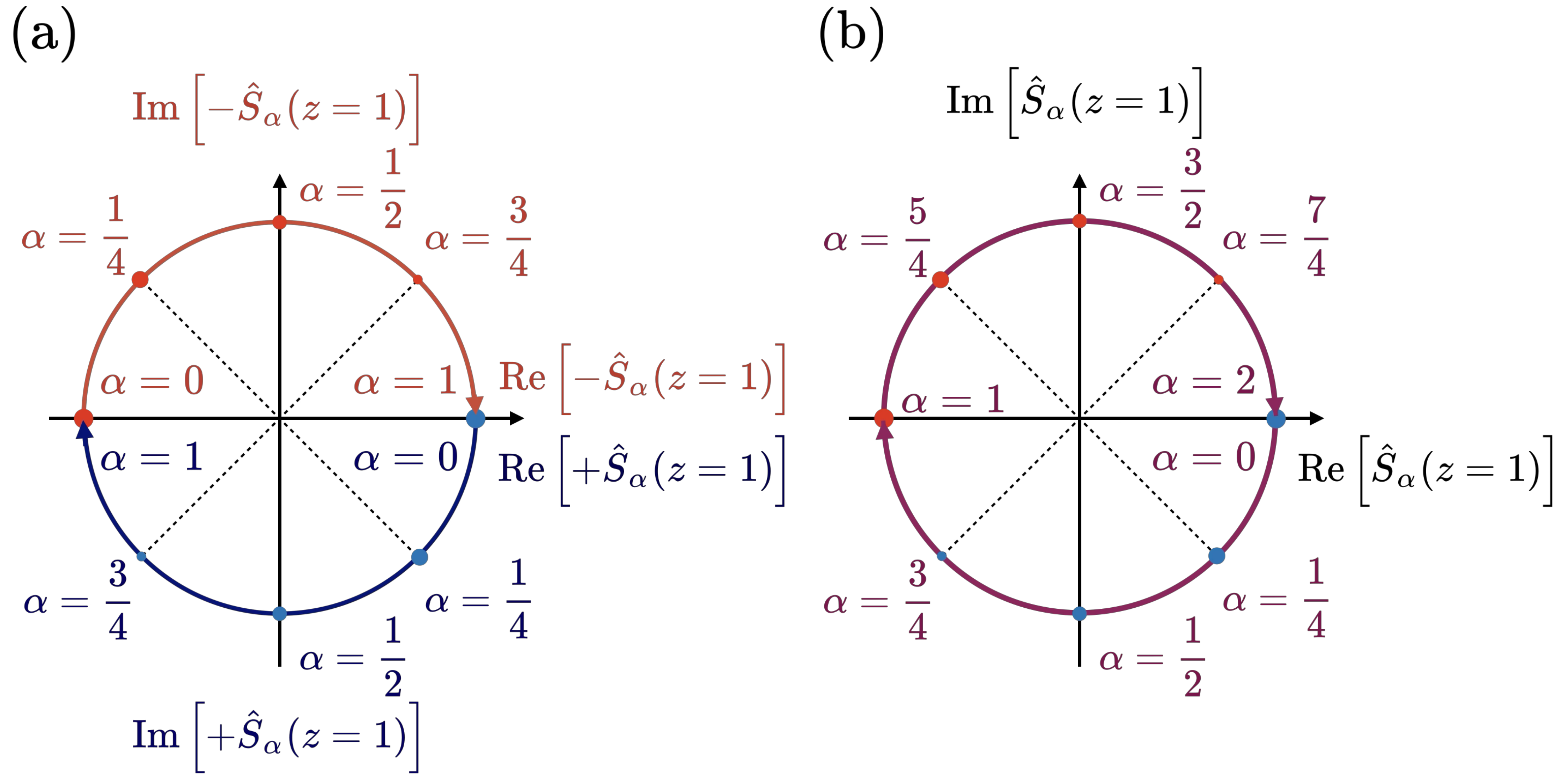}
    \caption{Illustration of exchange operator $\hat{S}_{\alpha}(z)$ for $z=1$ in the complex plane. 
    (a) Definition employed throughout this work. Our work distinguishes between  bosonic anyons ($+\hat{S}_\alpha(z)$ with $\alpha \in [0,1]$; lower half-circle in blue) and fermionic anyons ($-\hat{S}_\alpha(z)$ with $\alpha \in [0,1]$; upper half-circle in red).  (b) An alternative definition of the exchange operator that does not distinguish between bosonic anyons and fermionic anyons and would instead use $\hat{S}_{\alpha}(z)$ with  $\alpha \in [0,2]$.} 
    \label{fig_S_alpha_domain}
\end{figure}

We now define bosonic anyon exchange statistics for the relative wave function $\psi_{\alpha,+}(z)$ and fermionic anyon exchange statistics for the relative wave function  $\psi_{\alpha,-}(z)$
through~\cite{paper1} 
\begin{eqnarray}
\label{eq_def_anyonic_exchange}
\psi_{\alpha,\pm}(-z)= \pm\hat{S}_{\alpha}(z) \psi_{\alpha,\pm}(z),
\end{eqnarray}
where the “exchange operator” $\hat{S}_{\alpha}(z)$ is given by
\begin{eqnarray}
\label{eq_def_anyonic_exchange2}
\hat{S}_{\alpha}(z)= \exp \left[ -\iu \pi \alpha \sign(z) \right].
\end{eqnarray}
The exchange operator has the property $(\hat{S}^\dagger_{\alpha}(z))^{-1} = \hat{S}_{\alpha}(z)$. The sign-function is defined as~\cite{wang2024boson, paper1}
\begin{eqnarray}
\sign(z) = \left\{ \begin{array}{cl}
-1 & \mbox{for } z < 0 \\
\mbox{undefined } & \mbox{for } z=0\\
+1 & \mbox{for } z > 0 
\end{array} \right.
\end{eqnarray}
and $\sign^2(z)=1$ for all $z$~\cite{ValienteBFdualities}.  
 It is important to note that the 
anyonic exchange statistics 
introduced above distinguishes
between bosonic anyons and fermionic anyons, 
with
$\alpha$ 
going from $0$ to $1$ for both types of anyons~\cite{paper1}.
This is distinct from the literature~\cite{kundu1999exact, girardeau2006anyon, zinner2015strongly, wang2024boson}.
Our definition is
illustrated
in Fig.~\ref{fig_S_alpha_domain}(a),
where the operator $\hat{{S}}_{\alpha}(z)$, applicable to bosonic anyons, is presented by the blue half-circle in the lower part of the figure and the operator $-\hat{{S}}_{\alpha}(z)$, applicable to fermionic anyons, is presented by the red half-circle in the upper part of the figure (to make the figure, we set $z=1$).
 Alternatively, we could have defined one type of anyon [defining the exchange symmetry through  $\psi_{\alpha}(-z)=\hat{S}_{\alpha}(z) \psi_{\alpha}(z)$], with 
 $\alpha$ ranging from $0$ to $2$. This alternative definition, which is illustrated in Fig.~\ref{fig_S_alpha_domain}(b), is less appealing as it does not
 highlight the unique roles of the naturally occurring bosonic and fermionic exchange statistics (see also below).
It is worth noting that 
the exchange operator $\hat{S}_{\alpha}(z)$ has been defined in some papers with a plus sign as opposed to a minus sign in the exponent while considering only either the overall ``$+$" sign or the  overall ``$-$" sign on the right-hand side of Eq.~(\ref{eq_def_anyonic_exchange})~\cite{kundu1999exact, girardeau2006anyon,wang2024boson}.

We note that relative wave functions that possess bosonic anyon and fermionic anyon statistics can be obtained from bosonic and fermionic wave functions via~\cite{paper1}
\begin{eqnarray}
\label{eq_construction}
    \psi_{\alpha,\pm}(z)=\mathcal{N}(\alpha){\hat{S}}_{\alpha/2}^{\dagger}(z) \psi_{\pm}(z),
\end{eqnarray}
where ${\cal{N}}(\alpha)$ denotes a normalization factor that is defined in 
Eq.~(\ref{eq_normalization_anyon}).
The construction of anyonic wave functions 
given in Eq.~(\ref{eq_construction}) will be utilized in what follows. For later reference, it is also useful to rewrite $\hat{S}_{\alpha}(z)$ in terms of $\sin$ and $\cos$ functions,
\begin{eqnarray}
\hat{S}_{\alpha}(z)= \cos( \pi \alpha  )-\iu \sign(z) \sin ( \pi \alpha ) ,
\end{eqnarray}
where we used that $\cos$ is an even function and sin an odd function.

\subsection{Reference functions}

This section provides explicit expressions for the regular and irregular outside solutions
$f_k(z)$ and $g_k(z)$, respectively.
The regular and irregular outside solutions $f_{k;+}(z)$ and $g_{k;+}(z)$ for two identical bosons
are reported in the first line of Table~\ref{tab_reference_solutions}.
Both 
$f_{k;+}(z)$ and $g_{k;+}(z)$
obey the bosonic exchange statistics, i.e., they are even functions in $z$.
The outside solutions are chosen such that an infinitesimally attractive two-body potential
gives rise to a phase shift of $\pi/2 + 0^+$~\cite{kanjilal2004nondivergent, kanjilal2009pseudopotential}. Correspondingly, the 
even-parity scattering length $a_{+}$ [see Eq.~(\ref{eq_ascatt})] is equal to $+\infty$ for an infinitesimally attractive two-body potential,
reflecting the fact that an infinitesimal attraction in 1D leads to the formation of a shallow bound state~\cite{kanjilal2004nondivergent, kanjilal2009pseudopotential}.
 The regular and irregular outside solutions 
 $f_{k;-}(z)$ and $g_{k;-}(z)$ 
 for two identical fermions, which are odd functions in $z$,
are reported in the second line of Table~\ref{tab_reference_solutions}.
In this case, an infinitesimal attraction leads to an odd-parity scattering length $a_{-}$ that is small and negative, reflecting
the fact that a critical attraction is needed for a shallow bound state to be supported~\cite{kanjilal2004nondivergent, kanjilal2009pseudopotential}.

In analogy to the bosonic and fermionic outside solutions, the regular and irregular outside solutions for 
bosonic anyons and fermionic anyons are defined such that
they obey bosonic anyon exchange statistics [we denote the regular and irregular functions by $f_{k;\alpha,+}(z)$ and $g_{k;\alpha,+}(z)$,
respectively]
and fermionic anyon exchange statistics [we denote the regular and irregular functions by $f_{k;\alpha,-}(z)$ and $g_{k;\alpha,-}(z)$,
respectively].
In addition, we introduce an $\alpha$-dependent normalization factor ${\cal{N}}(\alpha)$,
\begin{eqnarray}
\label{eq_normalization_anyon}
{\cal{N}}(\alpha)=\frac{(1-\alpha) - \iu\alpha}{[(1-\alpha)^2 + \alpha^2]^{1/2}},
\end{eqnarray}
 such that
$f_{k;\alpha,\pm}(z)$ approaches   $f_{k;\pm}(z)$ and $f_{k;\mp}(z)$  for $\alpha=0$  and   $\alpha=1$, respectively.
Analogously, the irregular function 
$g_{k;\alpha,\pm}(z)$ approaches $g_{k;\pm}(z)$ 
and $g_{k;\mp}(z)$ 
for $\alpha=0$
and $\alpha=1$, respectively.
Since $|{\cal{N}}(\alpha)|^2=1$, ${\cal{N}}(\alpha)$ 
can be interpreted as  
an overall phase.
It can be checked 
that the expressions provided in the third and fourth rows of Table~\ref{tab_reference_solutions}
fulfill the exchange and normalization demands listed above.
Using Eq.~(\ref{eq_construction}),
$f_{k;\alpha,\pm}(z)$
and 
$g_{k;\alpha,\pm}(z)$
can be generated via
\begin{eqnarray}
f_{k;\alpha,\pm}(z)={\cal{N}}(\alpha) \hat{S}^{\dagger}_{\alpha/2}(z) f_{k;\pm}(z)
\end{eqnarray}
and
\begin{eqnarray}
g_{k;\alpha,\pm}(z)={\cal{N}}(\alpha) \hat{S}^{\dagger}_{\alpha/2}(z) g_{k;\pm}(z),
\end{eqnarray}
respectively.

 \begin{widetext}

 \begin{table}[h]
        \centering
        \begin{tabular}{l|l|l}
             \enspace 
             $s$ & Regular reference function $f_{k;s}(z)$  & Irregular reference function $g_{k;s}(z)$ \\ \hline
        $\quad +$ & $\mbox{sign}(z) \sin(kz)$ & $\cos(kz)$ \\
        $\quad -$ & $\sin(kz)$ & $\mbox{sign}(z)\cos(kz)$ \\
        $\alpha,+$ & $\mathcal{N}(\alpha)\left[\cos\left(\frac{\pi \alpha}{2}\right)\mbox{sign}(z) \sin(kz) + \iu \sin\left(\frac{\pi \alpha}{2}\right)\sin(kz)\right]$ & $\mathcal{N}(\alpha)\left[\cos\left(\frac{\pi \alpha}{2}\right)\cos(kz) + \iu \sin\left(\frac{\pi \alpha}{2}\right)\mbox{sign}(z)\cos(kz)\right]$ \\
        $\alpha,-$ & $\mathcal{N}(\alpha)\left[\cos\left(\frac{\pi \alpha}{2}\right)\sin(kz) + \iu \sin\left(\frac{\pi \alpha}{2}\right)\mbox{sign}(z)\sin(kz)\right]$ & $\mathcal{N}(\alpha)\left[\cos\left(\frac{\pi \alpha}{2}\right)\mbox{sign}(z)\cos(kz) + \iu \sin\left(\frac{\pi \alpha}{2}\right)\cos(kz) \right]$ \\   
        \end{tabular}
        \caption{Regular and irregular reference solutions for two identical particles with particle exchange statistics $s$. The subscript
        ``$s$'' can take the values ``$+$'' (boson),`` $-$'' (fermion), ``$\alpha,+$'' (bosonic anyon), and 
        ``$\alpha,-$'' (fermionic anyon).}
        \label{tab_reference_solutions}
\end{table}

It is instructive to reexpress the asymptotic solution $\psi_{k;\alpha,\pm}(z)$ [Eq.~(\ref{eq_outside}) with $f_k(z)$ and $g_k(z)$
replaced by  $f_{k;\alpha,\pm}(z)$ and $g_{k;\alpha,\pm}(z)$, respectively]
by rewriting the reference functions $f_{k;\alpha,\pm}(z)$ and $g_{k;\alpha,\pm}(z)$ in terms of
the even and odd reference functions $f_{k;\pm}(z)$ and $g_{k;\pm}(z)$,
\begin{eqnarray}
\label{eq_13}
\psi_{k;\alpha,\pm}(z)
\underset{|z| \ge z_0}{\longrightarrow}
\beta(k) {\cal{N}}(\alpha)
\left\{
\cos \left( \frac{\pi \alpha}{2} \right) \left[ f_{k;\pm}(z) +\tan(\delta_{\text{sc}}(k)) g_{k;\pm}(z) \right]
+
\iu
\sin \left( \frac{\pi \alpha}{2} \right) \left[ f_{k;\mp}(z) +\tan(\delta_{\text{sc}}(k)) g_{k;\mp}(z) \right]
\right\}.\nonumber \\
\end{eqnarray}
 \end{widetext}
 Equation~(\ref{eq_13}) shows that the K-matrix element  $\tan(\delta_{\text{sc}}(k))$, defined as quantifying---as usual~\cite{bransden2006physics}---the relative importance of the
 regular and irregular solutions with proper bosonic anyon and fermionic anyon exchange statistics,
    also quantifies the relative importance of the regular and irregular even parity functions
    and the relative importance of the regular and irregular odd parity functions.
    Equation~(\ref{eq_13}) will be used in the next section to identify a two-body Hamiltonian that is consistent
    with the asymptotic anyonic solutions introduced above.
    
\subsection{Anyonic interaction potential}     

Any wave function in 1D can be decomposed into its even and odd parts. Correspondingly, one
can  write a two-body interaction potential  $V(z)$ as a sum of two terms, namely $V_+(z)$ and $V_-(z)$:  $V_+(z)$ acts only on the
even part of the wave function and $V_-(z)$ acts only on the  
odd part of the wave function. This selectivity can be accomplished through the introduction of projectors onto the even and odd parity sectors.
According to Eq.~(\ref{eq_13}), the outside solution for bosonic anyons 
contains an even part [terms in the first pair of square brackets on the right hand side of Eq.~(\ref{eq_13})] and an odd part [terms in the second pair of square brackets on the right hand side of Eq.~(\ref{eq_13})]; the relative
importance of the regular and irregular functions in both the even and odd parts is given by $\tan(\delta_{\text{sc}}(k))$.
Similarly, the outside solution for fermionic anyons 
contains an odd part (terms in the first pair of square brackets) and an even part (terms in the second pair of square brackets); the relative
importance of the regular and irregular functions in both the odd and even parts is given by $\tan(\delta_{\text{sc}}(k))$.
These observations indicate that the two-body potential must be, to allow for eigenstates with anyonic exchange symmetry to be supported, such that $V_+(z)$ and $V_-(z)$ generate the same 
phase shift $\delta_{\text{sc}}(k)$.

In addition to the outside solution, the inside solution, i.e., the solution for $|z| < z_0$, must also obey the 
condition imposed by the anyonic exchange statistics.
In general, the even and odd inside solutions of finite-range potentials $V_+(z)$ and $V_-(z)$ that lead to the same phase shift will 
have functional forms that cannot be made to obey the anyonic exchange statistics, which requires a rather unique ``connection'' between the even and odd wave function parts.
This motivates the introduction of a zero-range interaction potential, for which the phase shift is accumulated at $z=0$, i.e., for which the outside
solution extends from $\infty$ to $0^+$ for positive $z$ and from $-\infty$ to $0^-$ for negative $z$.
Specifically, the potential 
\begin{eqnarray}
\label{eq_pot_pseudo}
    V_{\text{pseudo}}(z)=V_+(z)+V_-(z)
\end{eqnarray}
with
\begin{eqnarray}
V_+(z)= g_+(k) \delta(z)
\end{eqnarray}
and~\cite{bender2005exponentially}
\begin{eqnarray}
\label{eq_pot_odd}
V_-(z)= g_-(k) \frac{1}{z} \delta(z) \frac{\partial}{\partial z},
\end{eqnarray}
where
\begin{eqnarray}
g_+(k)=\frac{\hbar^2}{\mu} \frac{k}{\tan(\delta_{\text{sc}}(k))}
\end{eqnarray}
and
\begin{eqnarray}
g_-(k)=-\frac{\hbar^2}{\mu} \frac{\tan(\delta_{\text{sc}}(k))}{k},
\end{eqnarray}
supports solutions consistent with bosonic anyon and fermionic anyon exchange statistics.
The odd-parity pseudo-potential, Eq.~(\ref{eq_pot_odd}), can alternatively be written in terms of two derivatives, one that acts to the left and one that acts to the right~\cite{girardeau2003fermi, kanjilal2004nondivergent, grosse2004exact},
 \begin{eqnarray}
V_-(z)=g_-(k) \frac{\overleftarrow{\partial}}{\partial z} \delta(z) \frac{\overrightarrow{\partial}}{\partial z}.
\end{eqnarray}
Importantly, the constructed zero-range pseudo-potential also describes the scattering of two identical
bosons and of two identical fermions~\cite{paper1}. In the former case, $V_+(z)$ introduces an, in general, finite phase shift while $V_-(z)$ does not.
 In the latter case, $V_-(z)$ introduces an, in general, finite phase shift while $V_+(z)$ does not.

While it may be possible to construct a finite-range interaction potential that supports solutions consistent with the
anyonic exchange statistics, we are not aware of any such example. In practice, working in the low-energy limit, the range of the interactions
is small compared to the other length scales of the problem. This implies that a cold-atom realization of interacting anyons would violate the
anyonic exchange statistics requirements in the inner $|z|<z_0$ region, which constitutes a negligibly small region of the Hilbert space.
Consistent with the low-energy assumption, we use in the following  the energy-independent scattering length $a_{\text{sc}}$ [see Eq.~(\ref{eq_ascatt})]
to parametrize the interaction strengths, i.e., we use $k$-independent coupling strengths $g_+$ and $g_-$~\cite{kanjilal2004nondivergent},
\begin{eqnarray}
g_+ = -\frac{\hbar^2}{\mu a_{\text{sc}}}
\end{eqnarray}
and
\begin{eqnarray}
g_- = \frac{\hbar^2 a_{\text{sc}}}{\mu }.
\end{eqnarray}
It follows that
\begin{eqnarray}
    \label{eq_bafa_mapping_twobody}
    \psi_{\alpha, +}(z) = \sign(z)\psi_{\alpha, - }(z),
    \label{eq_BA_FA_mapping}
\end{eqnarray}
which is the two-particle version of the bosonic-anyon---fermionic-anyon mapping~\cite{paper1}. For $\alpha=0$, this reduces to the two-particle version of the celebrated Bose-Fermi mapping~\cite{girardeau2003fermi, girardeau2004effective, kanjilal2004nondivergent}, namely $\psi_+(z)=\sign(z)\psi_-(z)$.

The pseudo-potential $V_{\text{pseudo}}(z)$ can alternatively be formulated in terms of a short-range boundary condition, namely the 
logarithmic derivative of the relative wave function for $s$-particle exchange statistics  in the limits $z\rightarrow 0^+$
and $z\rightarrow 0^-$,
\begin{eqnarray}
\label{eq_logderivative}
    \frac{\frac{\partial \psi_{s}(z)}{\partial z}}{\psi_s(z)}\biggr\rvert_{|z| = 0^+} 
    =-\frac{1}{a_\mathrm{sc}}.
\end{eqnarray}
Equation~(\ref{eq_logderivative}) states
that the pseudo-potential governs the small-$|z|$ behavior of the relative wave function up to order $z$. The pseudo-potential does not impose any constraints on higher order terms like the $z^2$, $z^3$, etc. terms nor the normalization.
For later reference, we report the short-distance boundary conditions for two identical bosons, two identical fermions, two identical bosonic anyons, and two identical fermionic anyons explicitly as
\begin{align}
    \label{eq_BC_bosons}
    \psi_+(z) \underset{|z| \rightarrow 0^+}{\longrightarrow}& 1-\frac{|z|}{a_\mathrm{sc}},\\
    \label{eq_BC_fermions}
    \psi_-(z) \underset{|z| \rightarrow 0^+}{\longrightarrow}&
    \sign(z)-\frac{z}{a_\mathrm{sc}},\\
    \label{eq_BC_bosonicanyons}
    \psi_{\alpha, +}(z) \underset{|z| \rightarrow 0^+}{\longrightarrow} &\cos\left(\frac{\pi \alpha}{2} \right) \left(1-\frac{|z|}{a_\mathrm{sc}} \right) + \nonumber \\
    &\iu \sin \left(\frac{\pi \alpha}{2}\right) \left[\sign(z)-\frac{z}{a_\mathrm{sc}} \right],
\end{align}
and 
\begin{eqnarray}
    \label{eq_BC_fermionicanyons}
    \psi_{\alpha, -}(z) \underset{|z| \rightarrow 0^+}{\longrightarrow} \cos\left(\frac{\pi \alpha}{2}\right) \left[\sign(z)-\frac{z}{a_\mathrm{sc}} \right] + \nonumber \\
    \iu \sin \left(\frac{\pi \alpha}{2}\right) \left(z-\frac{|z|}{a_\mathrm{sc}} \right).
\end{eqnarray}

\subsection{Determination of bound state  via analytical continuation}
\label{sec_free_bound_anyons}
To determine the free-space bound state energy, we analytically continue the anyonic scattering solutions $\psi_{k; \alpha, \pm}(z)$, i.e., we replace
$k$  by $\iu \kappa$ in Eqs.~(\ref{eq_wavevector}) and
(\ref{eq_13}). 
Taking $\kappa \ge 0$, the bound state energy $E^{(\text{bd})}$ and the outside solution for bosonic anyons (fermionic anyons can be treated analogously) become
\begin{eqnarray}
E^{(\text{bd})} = -\frac{\hbar^2 \kappa^2}{2 \mu}
\end{eqnarray}
and 
\begin{eqnarray}
\label{eq_scatt_continued}
\psi^{(\text{bd})}_{\alpha, +}(z) \rightarrow \nonumber \\
\frac{\beta(k)}{2}\left[A_{k;\alpha, +}(z)\exp(-\kappa z)
+B_{k;\alpha, +}(z)\exp(\kappa z)
\right],
\end{eqnarray}
respectively, where 
\begin{eqnarray}
    A_{k;\alpha, +}(z) = 
    \mathcal{N}(\alpha)\exp\left[ \iu \frac{\sign(z) \pi \alpha}{2}  \right] \times \nonumber \\ 
    \left[-\iu\sign(z) +\tan(\delta_\mathrm{sc}(k)) \right],
    \nonumber \\
\end{eqnarray}
and
\begin{eqnarray}
    B_{k;\alpha, +}(z) = \mathcal{N}(\alpha)\exp\left[ \iu \frac{\sign(z) \pi \alpha}{2}  \right] \times \nonumber \\ 
    \left[\iu \sign(z) + \tan(\delta_\mathrm{sc}(k))\right]. \nonumber \\
\end{eqnarray}
Since Eq.~(\ref{eq_scatt_continued}) is valid for $|z| > z_0$, it should vanish as $z\rightarrow \pm \infty$.
For $z \rightarrow +\infty$, the $\exp(\kappa z)$ term blows up. This can be avoided by forcing $B_{k;\alpha, +}(z)$ to vanish.
Using that $\sign(z)=1$ for positive $z$, we  obtain
\begin{eqnarray}
\label{eq_pole}
\tan (\delta_\mathrm{sc}(k))=-\iu.
\end{eqnarray}
Similarly, 
for $z \rightarrow -\infty$, the $\exp(-\kappa z)$ term blows up. Forcing $A_{k;\alpha, +}(z)$ to go to zero yields the same bound state condition as before, namely  Eq.~(\ref{eq_pole}).
Recalling $k=\iu \kappa$,
Eq.~(\ref{eq_pole}) can be written as
\begin{eqnarray}
\kappa = - \frac{1}{\frac{\tan(\delta_\mathrm{sc}(k))}{k}}.
\end{eqnarray}
Since $\kappa \ge 0$, we see that a bound state only exists
if $-\tan(\delta_\mathrm{sc}(k))/k \ge 0$.
In the low-energy limit,
this implies that $a_{\text{sc}}$ must be positive. Neglecting the $k$-dependence of the scattering phase shift, the binding energy becomes
\begin{eqnarray}
\label{eq_energy_boundanyon}
E^{(\text{bd})} = - \frac{\hbar^2}{2 \mu (a_\mathrm{sc})^2}.
\end{eqnarray}
In terms of $a_\mathrm{sc}$, the normalized bosonic anyon bound state wavefunction for $|z|>z_0$ reads 
\begin{eqnarray}
\label{eq_bound_bosonicanyon}
\psi^{(\text{bd})}_{\kappa;\alpha, +}(z) = 
\mathcal{N}(\alpha)\frac{1}{\sqrt{a_\mathrm{sc}}}
\exp\left[ \iu \frac{\sign(z) \pi \alpha}{2}  \right]
\exp \left(- \frac{|z|}{a_\mathrm{sc}} \right).
  \nonumber \\
\end{eqnarray}
The bound state determined here via analytic continuation of the outside scattering solution
is, by construction, a bosonic anyon  eigenstate  of the relative Schr\"odinger equation with the pseudo-potential $V_{\text{pseudo}}(z)$.   
The fermionic anyon bound state solution can be found analogously (its energy is the same as that for two identical bosonic anyons). Alternatively, 
using the bosonic anyon--fermionic anyon mapping relation, Eq.~(\ref{eq_BA_FA_mapping}), we readily obtain the normalized fermionic anyon bound state wavefunction $\psi^{(\text{bd})}_{\alpha, -}(z)$,
\begin{eqnarray}
\label{eq_bound_fermionicanyon}
\psi^{(\text{bd})}_{\alpha, -}(z) = \sign(z)\psi^{(\text{bd})}_{\alpha, +}(z).
  \nonumber \\
\end{eqnarray}
We note that the bound state treatment breaks down when $a_{\text{sc}}$ approaches $0^+$. Physically, this corresponds to an extremely tightly bound state that cannot be captured by the low-energy theory developed in this work. Mathematically, the wave function would be non-zero at a single point only, making the limits $z \rightarrow 0^{\pm}$ ill-defined.

Recapitulating, this section determined the bound state energy and wave function for two identical bosonic anyons through analytic continuation of the scattering solution for two identical anyons. It is reassuring to see that the scattering formulation leads to the usual pole condition~\cite{bransden2006physics}, namely Eq.~(\ref{eq_pole}). We should note that the anyonic bound state wave functions could have alternatively been obtained by applying $\hat{S}_{\alpha/2}^{\dagger}(z)$ to the known bound state solutions for two identical bosons and two identical fermions [see Eq.~(\ref{eq_construction})].

\section{Off-diagonal observables}
\label{sec_fewbd_observables}
When the bosonic and fermionic $N$-particle systems interact through two-body zero-range pseudo-potentials that are characterized by the same scattering length, local or diagonal correlators such as the single-particle probability density and the static structure factor 
of these two 1D systems are the same~\cite{sekino2018comparative}.
Similarly, the two-body Tan contact ${\cal{C}}_2$, which enters as a multiplicative factor into the tail of the momentum distribution~\cite{tan2008energetics, tan2008large, tan2008generalized,sekino2018comparative,cui2016universal,paper1}, is a local observable that is insensitive to the particle statistics. Specifically, if $\Psi_s(z_1,z_2)$  denotes a two-particle wave function,
then the two-body contact ${\cal{C}}_2$,
\begin{eqnarray}
    \label{eq_twobody_contact}
    \mathcal{C}_{2} = 2\int^{\infty}_{-\infty} dz_1 |\Psi_s(z_1,z_1)|^2,
\end{eqnarray}
is the same for bosonic, fermionic, bosonic anyon, and fermionic anyon statistics (i.e., for $s=+$; $s=-$;  $s=\alpha,+$; and $s=\alpha,-$), provided the wavefunctions $\Psi_s(z_1,z_2)$ are related via the bosonic-anyon---fermionic-anyon mapping [see Eq.~(\ref{eq_bafa_mapping_twobody})]. The independence of ${\cal{C}}_2$ of the particle exchange statistics follows straightforwardly from the bosonic-anyon---fermionic-anyon mapping using $|\sign(z)|^2=1$ and $\hat{S}_{\alpha/2}^{\dagger}(z)\hat{S}_{\alpha/2}(z)=1$. While our work focuses on two-particle states, we note that the definition of the two-body contact extends to $N$-particle systems~\cite{paper1}. The three-body Tan contact ${\cal{C}}_3$, e.g., is used in Tables~\ref{tab_momentum_tail} and \ref{tab_momentum_tail_fermion}.

In contrast, off-diagonal elements of correlators such as the density matrix and momentum distribution of identical bosons and identical fermions  are not the same, even when their scattering lengths are equal (i.e., when the bosonic and fermionic eigenstates are related via the Bose-Fermi mapping): off-diagonal observables depend on the particle exchange statistics~\cite{bender2005exponentially, sekino2018comparative}. This implies that fingerprints of the chiral properties of bosonic anyon wavefunctions and fermionic anyon wavefunctions should be imprinted onto the off-diagonal elements of certain correlators. Section~\ref{sec_nonlocal} defines and reviews selected observables while Sec.~\ref{sec_nonlocal_symm}  
discusses, focusing on bosonic anyons and fermionic anyons,  their symmetry properties.

\subsection{Definitions and summary of select literature results}
\label{sec_nonlocal}
 
The one-body density matrix $\rho_s(z_1,z_1')$ is defined through~\cite{bender2005exponentially}
\begin{eqnarray}
\label{eq_obdm_general}
    \rho_s(z_1,z_1') = 2 \int^{L/2}_{-L/2} dz_2 \Psi_s^*(z_1',z_2)\Psi_s(z_1,z_2), 
\end{eqnarray}
where $L$ is the ``length'' of the box that the particles ``live'' in.
The normalization is chosen such that
\begin{eqnarray}
    \int^{L/2}_{-L/2}\:dz_1 \: \rho_s(z_1,z_1) = 2.
\end{eqnarray} 
The one-body density matrix $\rho_s(z_1,z_1')$ provides information about the coherence of the system. It can be interpreted as measuring how easy or hard it is to destroy the first particle at position $z_1$ and to subsequently re-create it at position $z_1'$. While the diagonal elements $\rho_s(z_1,z_1)$, which coincide with the ``standard" single-particle density, measure local correlations, the off-diagonals $\rho_s(z_1,z_1')$ with $z_1 \ne z_1'$ measure correlations that are sensitive to the exchange statistics.

The momentum distribution $n_s(k)$ is defined as the Fourier transform of the one-body density matrix~\cite{bender2005exponentially},
\begin{eqnarray}
    n_s(k) = \int^{L/2}_{-L/2} dz_1 \int^{L/2}_{-L/2} dz_1' e^{-\iu k (z_1-z_1')}\rho_s(z_1,z_1').
    \label{eq_momentum_correlation_1}
\end{eqnarray}
The normalization is chosen as such
\begin{eqnarray}
    \frac{1}{2 \pi} \int^{\infty}_{-\infty}\: dk \: n_s(k) = 2.
\end{eqnarray}
Alternatively, the momentum distribution can be written in terms of the absolute value square of the Fourier transform with respect to $z_1$ of the wavefunction $\Psi_s(z_1,z_2)$, 
\begin{eqnarray}
    n_s(k) = 
    2\int^{L/2}_{-L/2} dz_2  
    \left| \int^{L/2}_{-L/2} dz_1 \: e^{-\iu k z_1} \Psi_s(z_1,z_2)\right|^2.
    \nonumber \\
    \label{eq_momentum_correlation_2}
\end{eqnarray}
The momentum distribution is an observable that depends on the particle exchange statistics of the quantum state. For example, it has been shown  that the momentum distribution of infinitely-strongly interacting bosons and that of non-interacting fermions are distinct~\cite{olshanii1998atomic} even though their wave functions are related through the Bose-Fermi mapping~\cite{girardeau1960relationship}.
Similarly, the momentum distribution of infinitely-strongly interacting fermions and that of non-interacting bosons are distinct~\cite{bender2005exponentially}.

The large-$|k|$ tails of the momentum distributions $n_{\alpha,+}(k)$ and $n_{\alpha,-}(k)$ for $N$ identical 1D bosonic anyons and $N$ identical 1D fermionic anyons, both interacting with two-body zero-range interactions with scattering length $a_{\text{sc}}$, were analyzed in our earlier work~\cite{paper1}.
Tables~\ref{tab_momentum_tail}  and
\ref{tab_momentum_tail_fermion} summarize the limiting behaviors for $N$ identical 1D bosonic anyons and $N$ identical 1D fermionic anyons, respectively.
For $0<|a_{\text{sc}}|< \infty$ and
$0<\alpha<1$, the $k^{-2}$ and $k^{-3}$ contributions to the tails are universal, i.e., fully determined by the two- and three-body Tan contacts ${\cal{C}}_2$ and ${\cal{C}}_3$ (which are local observables), the statistical parameter $\alpha$, and the scattering length $a_{\text{sc}}$~\cite{paper1}. The $k^{-4}$ contributions to the tails are, as will be shown in Sec.~\ref{sec_examples}, not fully universal as they contain a term that depends on the details of the state under consideration. An example is the lowest energy state of identical anyons in free space (see Sec.~\ref{sec_spdm_corr_md}) and under harmonic confinement (see Sec.~\ref{sec_ho_anyons}). 

The existence of universal and non-universal contributions to the momentum distribution tails can be understood as follows.  
Quite generally, algebraic powerlaw contributions to the momentum tail of the form $k^{-l}$ arise due to the discontinuities of the 
  wavefunction or its derivatives when two or more particles sit on top of each other~\cite{olshanii2003short, cui2016universal,sekino2018comparative, paper1}. The ratio between the wave function and its derivative when two particles sit on top of each other [i.e., the boundary condition enforced by the zero-range pseudo-potential and particle exchange statistics, see Eqs.~(\ref{eq_BC_bosons})-(\ref{eq_BC_fermionicanyons})], sets the leading-order contribution(s) to the tail of the momentum distribution.
  For bosonic and fermionic anyons, our earlier work~\cite{paper1} showed that the $k^{-2}$ and $k^{-3}$ contributions to the tails are universal. Intriguingly, the $k^{-3}$ term was shown to contain a contribution that arises due to discontinuities that involve three particles sitting on top of each other.  To fully determine the $k^{-4}$ contribution for the two-particle system, one needs to know the $z^2$ behavior of the wave function in the limit that the interparticle distance goes to zero  ($|z| \rightarrow 0$). Since this limiting behavior is not enforced by the two-body pseudo-potential $V_{\text{pseudo}}(z)$, the $k^{-4}$ tail is, in general, not fully universal (it contains universal and non-universal contributions). 
   For $N$ identical bosons with zero-range interactions characterized by a finite two-body scattering length, the $k^{-2}$ and $k^{-3}$ contributions are absent as is the non-universal $k^{-4}$ contribution. As a consequence, for $N$ identical bosons the leading-order contribution to the tail is proportional to $k^{-4}$ (this term is fully universal~\cite{olshanii2003short, sekino2018comparative, pactu2017universal}).
   For $N$ identical fermions,  the $k^{-2}$ contribution to the tail is universal~\cite{grosse2004exact,girardeau2003fermi, girardeau2004spinorFB, cui2016universal} while the next higher-order contribution, namely the $k^{-4}$ term, contains non-universal contributions.  

 \begin{widetext}

 \begin{table}[h]
        \begin{tabular}{c|c|c|c}
             \enspace   & $\alpha = 0$  & $0<\alpha<1$  & $\alpha = 1$\\ \hline
        $|a_\mathrm{sc}| = 0$ & $\lim_{|a_{\text{sc}}|=0}\frac{4\mathcal{C}_{2}}{a_{\text{sc}}^2 k^4}$\,\footnote{In the $|a_{\text{sc}}|\rightarrow 0$ limit, ${\cal{C}}_2$ is directly proportional to $a_{\text{sc}}^2$. It follows that the leading-order contribution to the tail is independent of $a_{\text{sc}}$ and directly proportional to $k^{-4}$.} & $\lim_{|a_{\text{sc}}|=0}\frac{4 {\mathcal{C}}_{2}}{a_{\text{sc}}^2k^4}\cos^2\left(\frac{\pi \alpha}{2} \right)\,^{\color{red}\mathrm{a}}$  & no algebraic tail \\
        $0<|a_\mathrm{sc}|<\infty$ & $\frac{4\mathcal{C}_{2}}{a^2_\mathrm{sc} k^4}$ & $\frac{4\mathcal{C}_{2}}{k^2}\sin^2\left(\frac{\pi \alpha}{2} \right) + \frac{4\mathcal{C}_{2}}{a_\mathrm{sc}k^3}\sin\left(\pi \alpha\right) + \frac{8\mathcal{C}_3}{k^{3}}\sin^2\left(\frac{\pi \alpha}{2}\right)\sin\left(\pi \alpha\right)$ & $\frac{4\mathcal{C}_{2}}{k^2}$ \\
        $|a_\mathrm{sc}|^{-1} = 0$ & no algebraic tail & $\frac{4\mathcal{C}_{2}}{k^2}\sin^2\left(\frac{\pi \alpha}{2} \right)$ & $\frac{4\mathcal{C}_{2}}{k^2}$ \\
        
        \end{tabular}
        \caption{Summary of the behavior of $\lim_{|k|\rightarrow\infty} n_{\alpha, +}(k)$ for $N$ identical bosonic anyons. Sub-leading terms, such as the $k^{-4}$ contribution for $0<|a_{\text{sc}}|<\infty$ and $0<\alpha<1$, contain universal and non-universal contributions that are not included in this table (see Sec.~\ref{sec_examples} for details). The limiting statistics of $\alpha=0$ (identical bosons) and $\alpha=1$ (identical fermions) are reported in the second and fourth columns. Since the wave functions of non-interacting bosons ($|a_{\text{sc}}|^{-1}=0$) and non-interacting fermions ($|a_{\text{sc}}|=0$) do not possess any discontinuities, their momentum distributions do not possess algebraic tails. The table entries are taken from Refs.~\cite{olshanii2003short, cui2016universal, paper1}.}
        \label{tab_momentum_tail}
\end{table} 

 \begin{table}[h]
        \begin{tabular}{c|c|c|c}
             \enspace   & $\alpha = 0$  & $0<\alpha<1$  & $\alpha = 1$\\ \hline
        $|a_\mathrm{sc}| = 0$ & no algebraic tail & $\lim_{|a_{\text{sc}}|=0}\frac{4 {\mathcal{C}}_{2}}{a_{\text{sc}}^2k^4}\sin^2\left(\frac{\pi \alpha}{2} \right)\,^{\color{red}\mathrm{a}}$  & $\lim_{|a_{\text{sc}}|=0}\frac{4\mathcal{C}_{2}}{a_{\text{sc}}^2 k^4}$\,\footnote{In the $|a_{\text{sc}}|\rightarrow 0$ limit, ${\cal{C}}_2$ is directly proportional to $a_{\text{sc}}^2$. It follows that the leading-order contribution to the tail is independent of $a_{\text{sc}}$ and directly proportional to $k^{-4}$.} \\
        $0<|a_\mathrm{sc}|<\infty$ & $\frac{4\mathcal{C}_{2}}{k^2}$ & $\frac{4\mathcal{C}_{2}}{k^2}\cos^2\left(\frac{\pi \alpha}{2} \right) - \frac{4\mathcal{C}_{2}}{a_\mathrm{sc}k^3}\sin\left(\pi \alpha\right) - \frac{8\mathcal{C}_3}{k^{3}}\cos^2\left(\frac{\pi \alpha}{2}\right)\sin\left(\pi \alpha\right)$ &  $\frac{4\mathcal{C}_{2}}{a^2_\mathrm{sc} k^4}$ \\
        $|a_\mathrm{sc}|^{-1} = 0$ & $\frac{4\mathcal{C}_{2}}{k^2}$ & $\frac{4\mathcal{C}_{2}}{k^2}\sin^2\left(\frac{\pi \alpha}{2} \right)$ & no algebraic tail \\
        
        \end{tabular}
        \caption{Summary of the behavior of $\lim_{|k|\rightarrow\infty} n_{\alpha, -}(k)$ for $N$ identical fermionic anyons. The table lists the universal algebraic contributions to the tail of the momentum distribution. Sub-leading terms, such as the $k^{-4}$ contribution for $0<|a_{\text{sc}}|<\infty$ and $0<\alpha<1$, contain universal and non-universal contributions that are not included in this table (see Sec.~\ref{sec_examples} for details). The limiting statistics of $\alpha=0$ (identical fermions) and $\alpha=1$ (identical bosons) are reported in the second and fourth columns. Since the wave functions of non-interacting fermions ($|a_{\text{sc}}|=0$) and non-interacting bosons ($|a_{\text{sc}}|^{-1}=0$) do not possess any discontinuities, their momentum distributions do not possess algebraic tails. The table entries are taken from Refs.~\cite{olshanii2003short, cui2016universal, paper1}.}
        \label{tab_momentum_tail_fermion}
\end{table} 
\end{widetext}
\subsection{Symmetry properties}
\label{sec_nonlocal_symm}
As discussed earlier, the bosonic-anyon---fermionic-anyon mapping relates the wave function of two identical bosonic anyons to that of two identical fermionic anyons, provided they are solutions to the zero-range Hamiltonian with the same scattering length. This section leverages this mapping  to derive relationships between the density matrices and the momentum distributions of bosonic anyons and fermionic anyons.  For the discussion that follows, it is useful to rewrite the sign-function, which enters into the $\hat{A}(z)$ and $\hat{S}_{\alpha/2}^{\dagger}(z)$ operators, in terms of $\hat{S}_{\alpha}(z)$ and its adjoint, 
\begin{eqnarray}
    \label{eq_trick_sign}
    \hat{A}(z) = \sign(z) =  \pm\iu \hat{S}_{\pm 1/2}(z) 
    =\mp\iu \hat{S}^\dagger_{\pm 1/2}(z).
\end{eqnarray}
Using $\Psi_+(z_1,z_2)=-\Psi_-(z_1,z_2)$ to reexpress the bosonic-anyon---fermionic-anyon mapping [see Eq.~(\ref{eq_bafa_mapping_twobody})], we find
\begin{eqnarray}
    \label{eq_relate_nonphys_ba1} \Psi_{\alpha,+}(z_1,z_2)=\hat{S}_{\alpha/2}^{\dagger}(z)  \sign(z)\Psi_-(z_1,z_2)
    \end{eqnarray}
    and, using Eq.~(\ref{eq_trick_sign}),
\begin{eqnarray}
    \label{eq_relate_nonphys_ba} \Psi_{\alpha,+}(z_1,z_2)=
    \mp \iu \Psi_{-,\alpha\pm1}(z_1,z_2).
\end{eqnarray}
Similarly, we find
\begin{eqnarray}
\label{eq_relate_nonphys_fa1}
    \Psi_{\alpha,-}(z_1,z_2)=\hat{S}_{\alpha/2}^{\dagger}(z)  \sign(z)\Psi_+(z_1,z_2)
\end{eqnarray}
and 
\begin{eqnarray}
\label{eq_relate_nonphys_fa}
    \Psi_{\alpha,-}(z_1,z_2)= \mp \iu \Psi_{+,\alpha\pm1}(z_1,z_2).
\end{eqnarray}
Equations~(\ref{eq_relate_nonphys_ba}) and (\ref{eq_relate_nonphys_fa}) establish a ``formal'' connection between the wave functions of bosonic anyons and those of fermionic anyons.
This formal connection is non-physical since the ranges of the ``effective statistical parameters'' $\alpha+1$ and $\alpha-1$ lie outside of the physical domain, which covers the values zero to one. Nevertheless, the continuation into the non-physical domain proves useful in some calculations. The formal connections readily yield
\begin{eqnarray}
    \rho_{\alpha,+}(z_1,z_1')=\rho_{\alpha \pm 1,-}(z_1,z_1'),
\end{eqnarray}
\begin{eqnarray}
    \rho_{\alpha,-}(z_1,z_1')=\rho_{\alpha \pm 1,+}(z_1,z_1'),
\end{eqnarray}
\begin{eqnarray}
    \label{eq_momentum_symm1}
    n_{\alpha,+}(k)=n_{\alpha \pm 1,-}(k),
\end{eqnarray}
and 
\begin{eqnarray}
\label{eq_momentum_symm2}
    n_{\alpha,-}(k)=n_{\alpha \pm 1,+}(k).
\end{eqnarray}
Equation~(\ref{eq_momentum_symm2}), e.g., was utilized in the supplemental material of Ref.~\cite{paper1} to determine the tail of $n_{\alpha,-}(k)$ in the physical domain.

Under the assumption that $\Psi_+(z_1,z_2)$ and $\Psi_-(z_1,z_2)$ are purely real functions, one can derive relationships between observables of bosonic anyons and fermionic anyons, with all quantities defined in the  physical domain. Since $\Psi_+(z_1,z_2)$ and $\Psi_-(z_1,z_2)$ can always chosen to be purely real, the relationships are general.
Taking the complex conjugates of Eqs.~(\ref{eq_relate_nonphys_ba}) and (\ref{eq_relate_nonphys_fa}) and rearranging the subsequent expressions, we find (see Appendix~\ref{app_derivation_some_symm} for details)
\begin{eqnarray}
    \label{eq_BA_FA_mirrorWF_1}
    \Psi_{\alpha,+}(z_1,z_2)=\iu \left[\Psi_{1-\alpha,-}(z_1,z_2)\right]^*
\end{eqnarray}
and
\begin{eqnarray}
    \label{eq_BA_FA_mirrorWF_2}
    \Psi_{\alpha,-}(z_1,z_2)= \iu \left[\Psi_{1-\alpha,+}(z_1,z_2)\right]^* .
\end{eqnarray}
Inserting these relationships into the definitions of $\rho_{\alpha,\pm}(z_1,z_1')$ and $n_{\alpha,\pm}(k)$, we find
\begin{eqnarray}
\label{eq_obdm_real}
    \mbox{Re} \left[ \rho_{\alpha, \pm} (z_1,z_1')\right]= \mbox{Re} \left[\rho_{1-\alpha, \mp} (z_1,z_1') \right],
\end{eqnarray}
\begin{eqnarray}
\label{eq_obdm_imag}
    \mbox{Im} \left[ \rho_{\alpha, \pm} (z_1,z_1')\right]= -\mbox{Im} \left[\rho_{1-\alpha, \mp} (z_1,z_1') \right],
\end{eqnarray}
and
\begin{eqnarray}
\label{eq_momentum_symm3}
    n_{\alpha,\pm}(k) = n_{1-\alpha,\mp}(-k).
\end{eqnarray}
Even though our derivation considered the two-particle case, it can be fairly straightforwardly extended to $N$ identical particles. It follows that
Eqs.~(\ref{eq_obdm_real})-(\ref{eq_momentum_symm3}) are applicable to any $N$.

\section{Examples}
\label{sec_examples}
\subsection{Two-anyon bound state in free space}
\label{sec_spdm_corr_md} 
\begin{figure*}[t]
    \centering
    \includegraphics[width=0.95\textwidth]{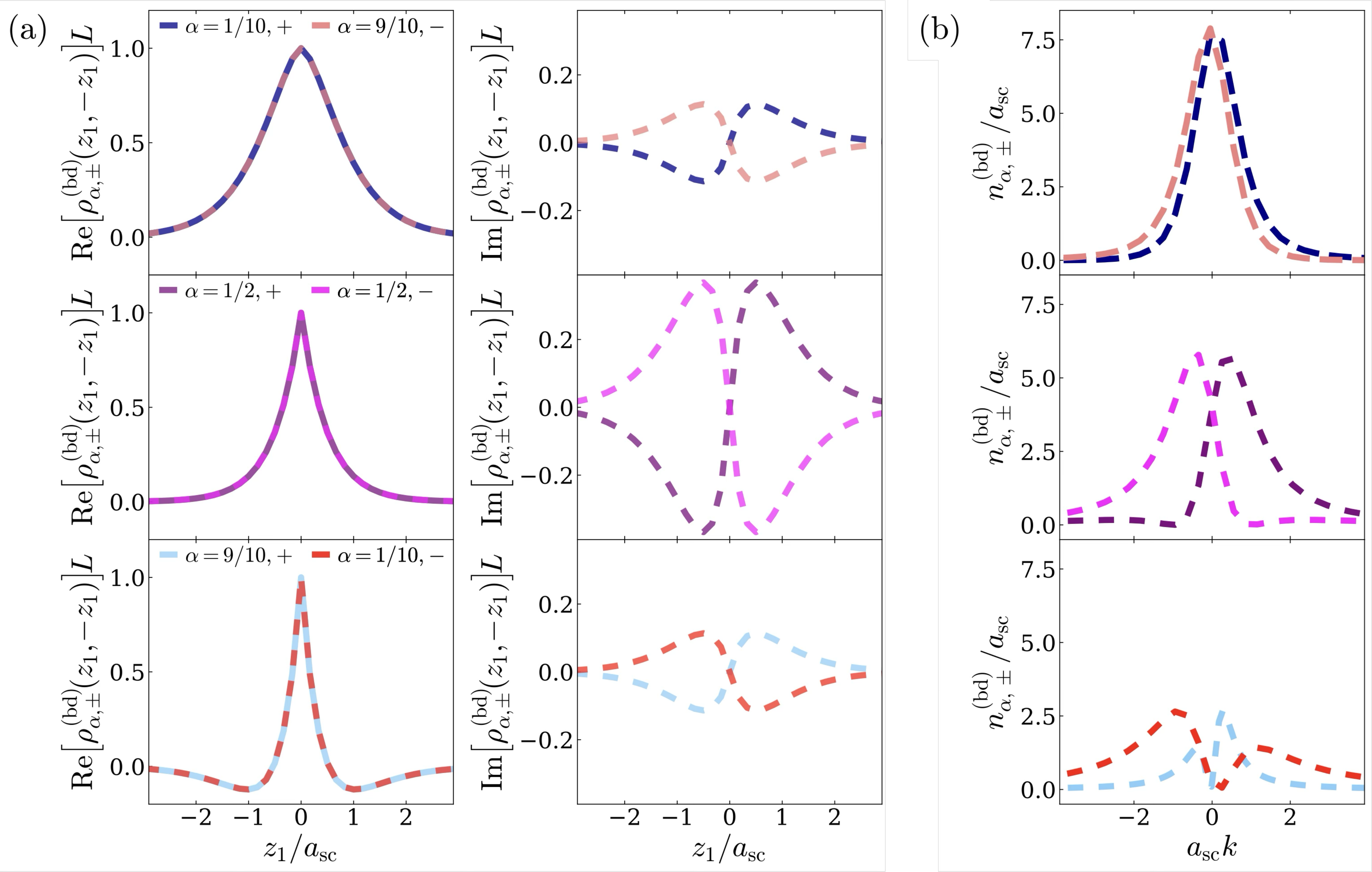}
    \caption{Properties of off-diagonal elements of anyonic bound state in free space. (a) The first and second columns show, respectively, the real and imaginary parts of the one-body density matrix $\rho_{\alpha, \pm}^{(\text{bd})}(z_1,-z_1)$  for $z_2=-z_1$ as a function of $z_1/a_{\text{sc}}$.
  (b) Momentum distribution $n_{\alpha, \pm}^{(\text{bd})}(k)$ as a function of  $a_{\mathrm{sc}} k$. Motivated by the symmetry properties derived in Sec.~\ref{sec_nonlocal_symm}, each row  compares bosonic anyons with statistical parameter $\alpha$ and fermionic anyons with statistical parameter $1-\alpha$.}
    \label{fig_SPDM_any}
\end{figure*}

This section discusses the one-body density matrix and the momentum distribution for the bound state of two bosonic anyons and two fermionic anyons. 
Setting the center-of-mass momentum to zero, the center-of-mass wavefunction reduces to $\sqrt{1/L}$ so that we have $\Psi^{(\text{bd})}_{\alpha, \pm}(z_1,z_2) = L^{-1/2}\psi^{(\text{bd})}_{\alpha, \pm}(z)$.
Inserting this expression into Eq.~(\ref{eq_obdm_general}) and breaking the integral over $z_2$ into two domains, namely $z_2>z_1$ ($z<0$) and $z_2<z_1$ ($z>0$), we find
\begin{eqnarray}
    \label{eq_SPDM_free_bosonicanyon}
    \rho^{(\text{bd})}_{\alpha, \pm} (z_1,z_1') =
    \frac{1}{L}\exp\left(-\frac{|z_1-z_1'|}{a_\mathrm{sc}}\right) \times \nonumber\\
    \left(1 \pm \exp\left[\iu \alpha \pi \sign(z_1-z_1')\right]\frac{|z_1 - z_1'|}{a_{\mathrm{sc}}}\right).
\end{eqnarray}
Equation~(\ref{eq_SPDM_free_bosonicanyon})  confirms the validity of Eqs.~(\ref{eq_obdm_real}) and (\ref{eq_obdm_imag}) for the specific example of two bound anyons in free space.
Figure~\ref{fig_SPDM_any} 
shows the real and imaginary parts of $\rho^{(\text{bd})}_{\alpha, \pm} (z_1,-z_1)L$ as a function of $z_1/a_{\mathrm{sc}}$ for various  $\alpha$. The plots provide a nice visualization of Eqs.~(\ref{eq_obdm_real}) and (\ref{eq_obdm_imag}).
The width of the central peak of the real part of $\rho^{(\text{bd})}_{\alpha, +} (z_1,-z_1)$, measured in units of $a_{\mathrm{sc}}$, decreases with increasing $\alpha$.
It is broad for $\alpha=0$ (bosons) and narrow for $\alpha=1$ (fermions).
Similarly, the width of the central peak of the real part of $\rho^{(\text{bd})}_{\alpha, -} (z_1,-z_1)$ decreases with decreasing $\alpha$.
It is broad for $\alpha=1$ (fermions) and narrow for $\alpha=0$ (bosons). It can be seen that $\mbox{Re}[\rho^{(\text{bd})}_{\alpha, \pm} (z_1,-z_1)]$
possesses a discontinuous derivative at $z_1=0$.
The imaginary part of $\rho^{(\text{bd})}_{\alpha, \pm} (z_1,-z_1)$ is
zero for $\alpha=0$ and $1$ (not shown in Fig.~\ref{fig_SPDM_any}) and, in general, non-zero for $0<\alpha<1$ for both bosonic anyons and fermionic anyons. For fixed $\alpha$,  $\mbox{Im}[\rho^{(\text{bd})}_{\alpha, \pm} (z_1,-z_1)]$ is an odd function in $z_1$. The global minimum and global maximum of $\mbox{Im}[\rho^{(\text{bd})}_{\alpha, \pm} (z_1,-z_1)]$ increase monotonically as $\alpha$ increases from $0$ to $1/2$ and then decrease monotonically as $\alpha$ increases from $1/2$ to $1$. For fixed $\alpha$, the $z_1/a_{\mathrm{sc}}$ value at which $\mbox{Im}[\rho^{(\text{bd})}_{\alpha, +} (z_1,-z_1)]$ is maximal is the value at which $\mbox{Im}[\rho^{(\text{bd})}_{\alpha, -} (z_1,-z_1)]$ is minimal, and vice versa.
 Since the momentum distribution can be expressed in terms of the Fourier transform of $\rho^{(\text{bd})}_{\alpha, \pm} (z_1,-z_1')$, the properties of the one-body density matrix get imprinted onto the momentum distribution.

The momentum distributions for two bound bosonic anyons and two bound fermionic anyons were reported in the supplemental material of Ref.~\cite{paper1}. We reproduce them here for completeness:
\begin{eqnarray}
\label{eq_bosonicany_bound_md_freespace}
    n^{(\text{bd})}_{\alpha,+}(k) = 8a_{\mathrm{sc}}\frac{\left[\cos(\frac{\pi \alpha}{2} ) + a_{\mathrm{sc}}k\sin(\frac{\pi \alpha}{2})\right]^2}{(1+a_{\mathrm{sc}}^2k^2)^2}
\end{eqnarray}
and
\begin{eqnarray}
\label{eq_fermionicany_bound_md_freespace}
    n^{(\text{bd})}_{\alpha,-}(k) = 8a_{\mathrm{sc}}\frac{\left[\sin(\frac{\pi \alpha}{2} ) - a_{\mathrm{sc}}k\cos(\frac{\pi \alpha}{2})\right]^2}{(1+a_{\mathrm{sc}}^2k^2)^2}.
\end{eqnarray}
In determining $n^{(\text{bd})}_{\alpha,-}(k)$, $L$ was taken to $\infty$. 
It can be checked readily that 
these momentum distributions obey Eqs.~(\ref{eq_momentum_symm1}), (\ref{eq_momentum_symm2}), and (\ref{eq_momentum_symm3}). 
Figure~\ref{fig_SPDM_any}(b) shows the scaled momentum distributions
$n_{\alpha,\pm}^{(\text{bd})}(k)/a_{\mathrm{sc}}$ as a function of $a_{\mathrm{sc}}k$ for various $\alpha$. 
It can be seen that the momentum distributions of interacting anyons are, for $0<\alpha<1$, not symmetric with respect to $k=0$: the momentum distributions are, as a consequence of the broken chiral symmetry, skewed. Inspection of Eqs.~(\ref{eq_bosonicany_bound_md_freespace}) and (\ref{eq_fermionicany_bound_md_freespace})
shows that the asymmetry comes from the $k$-dependence of the numerator.  Multiplying out the square in the numerator of Eqs.~(\ref{eq_bosonicany_bound_md_freespace}) and (\ref{eq_fermionicany_bound_md_freespace}), one sees that the cross term is linear in $k$. The cross term vanishes for $\alpha=0$ and $\alpha=1$, i.e., for bosons and fermions.

The positions of the extrema of the momentum distribution and the value of the momentum distribution at the extrema may
be used to quantify the chirality of the anyonic system.
The $k$ value at which $n_{\alpha, +}^{(\text{bd})}(k)$ takes on its global maximum is denoted by $k_{+,1}$ while that at which  $n_{\alpha, +}^{(\text{bd})}(k)$ takes on a local maximum is denoted by $k_{+,2}$ (see Table~\ref{table_momdist_peaks}).
The value of $k_{+,1}$   increases  monotonically from $0$ to $1/a_{\mathrm{sc}}$ as  $\alpha $ increases from $0$ to $1$. The value of $k_{+,2}$ increases monotonically from $-\infty$ to $-1/a_{\mathrm{sc}}$ as $\alpha$ increases from $0$ to $1$. 
For $\alpha = 0$ (two identical bosons), one finds $n_{0,+}^{(\text{bd})}(k_{+,1})=8a_\mathrm{sc}$ and $n_{0,+}^{(\text{bd})}(k_{+,2})=0$. 
For $\alpha = 1$ (two identical fermions), one finds $n_{1,+}^{(\text{bd})}(k_{+,1})=n_{1,+}^{(\text{bd})}(k_{+,2})=2a_\mathrm{sc}$.
Similarly, the $k$ value at which $n_{\alpha, -}^{(\text{bd})}(k)$ takes on its global maximum is denoted by $k_{-,1}$ while that at which  $n_{\alpha, -}^{(\text{bd})}(k)$ takes on a local maximum is denoted by $k_{-,2}$ (see Table~\ref{table_momdist_peaks}).
The value of $k_{-,1}$   increases  monotonically from $-1/a_{\mathrm{sc}}$ to $0$ as  $\alpha $ increases from $0$ to $1$. The value of $k_{+,2}$ increases monotonically from $1/a_{\mathrm{sc}}$ to $\infty$ as $\alpha$ increases from $0$ to $1$. 
For $\alpha = 0$ (two identical fermions), one finds $n_{0,-}^{(\text{bd})}(k_{-,1})=n_{0,-}^{(\text{bd})}(k_{-,2})=2a_\mathrm{sc}$. 
For $\alpha = 1$ (two identical bosons), one finds $n_{1,-}^{(\text{bd})}(k_{-,1})=8a_\mathrm{sc}$ and $n_{1,-}^{(\text{bd})}(k_{-,2})=0$.

 \begin{table}[h!]
\begin{center}
\begin{tabular}{ r |c}
$k_{+,1}$ & $\tan(\pi \alpha/4)/a_\mathrm{sc}$\\
$n_{\alpha,+}^{(\text{bd})}(k_{+,1})$ & $8a_\mathrm{sc}\cos^4(\pi \alpha/4)$\\
$k_{+,2}$ & $-\cot(\pi \alpha/4)/a_\mathrm{sc}$\\
$n_{\alpha,+}^{(\text{bd})}(k_{+,2})$ & $8a_\mathrm{sc}\sin^4(\pi \alpha/4)$\\ \hline
$k_{-,1}$ & $\left[\tan(\pi\alpha/2) - \sec(\pi\alpha/2)\right]/a_\mathrm{sc}$\\
$n_{\alpha,-}^{(\text{bd})}(k_{-,1})$ & $2 a_\mathrm{sc}\left[\cos(\pi\alpha/4)+\sin(\pi\alpha/4)\right]^4$\\
$k_{-,2}$ & $\left[\tan(\pi\alpha/2) + \sec(\pi\alpha/2)\right]/a_\mathrm{sc}$\\
$n_{\alpha,-}^{(\text{bd})}(k_{-,2})$ & $a_\mathrm{sc}\left[3 - \cos(\alpha \pi) - 4\sin(\pi\alpha/2)\right] $
\label{table_momdist_peaks}
\end{tabular}
\caption{Characterization of extrema of the momentum distributions 
$n_{\alpha, \pm}^{(\text{bd})}(k)$ for two identical bound anyons in free space.
The values of $k_{\pm,j}$ ($j=1$ or $2$), where the momentum distributions take on extrema, and
$n_{\alpha, \pm}^{(\text{bd})}(k_{\pm,j})$ are reported. 
}
\end{center}
\end{table}
 The analytic momentum distribution expressions for the anyonic dimers, Eqs.~(\ref{eq_bosonicany_bound_md_freespace}) and (\ref{eq_fermionicany_bound_md_freespace}), allow for the asymptotic expressions of the tail of the momentum distribution to be confirmed explicitly. 
  Expanding 
 the momentum distribution for bosonic anyons  given in Eq.~(\ref{eq_bosonicany_bound_md_freespace}), we find  
\begin{eqnarray}
\label{eq_free_tail_BA}
    \lim_{|k| \rightarrow \infty} n^{(\text{bd})}_{\alpha,+}(k) =
       \frac{4  {\cal{C}}^{(\text{bd})}_2}{k^2}  \sin^2 \left( \frac{\pi \alpha}{2} \right)
       + \frac{4  {\cal{C}}^{(\text{bd})}_2}{a_{\mathrm{sc}}k^3}
         \sin \left( \pi \alpha\right)  
        \nonumber \\
        + \frac{4  {\cal{C}}^{(\text{bd})}_2}{a^2_{\mathrm{sc}}k^4} \left[\cos^2 \left( \frac{\pi \alpha}{2} \right) - 2\sin^2 \left( \frac{\pi \alpha}{2} \right) \right]+{\cal{O}}(k^{-5}),\nonumber \\
\end{eqnarray}
where $\mathcal{C}^{(\text{bd})}_2 = 2/a_{\mathrm{sc}}$ is the two-body Tan contact for the bound state in free space~\cite{tan2008energetics, tan2008large, barth2011tan}. In Eq.~(\ref{eq_free_tail_BA}), we added the superscript ``$(\text{bd})$'' to distinguish this two-body contact from that for other states considered in this paper.
As mentioned earlier, the two-body contact is independent of the particle statistics, i.e., it can be calculated for the bosonic, fermionic, bosonic anyon, or fermionic anyon bound state wave functions: they all yield the same result.
In writing Eq.~(\ref{eq_free_tail_BA}), 
we used that the momentum tail is known to be proportional to ${\cal{C}}_2$; specifically, terms that scale with $({\cal{C}}_2)^2$ or
$({\cal{C}}_2)^3$ should not appear. 

For $\alpha=0$, Eq.~(\ref{eq_free_tail_BA}) reproduces the known $k^{-4}$ tail for identical bosons. 
 For $\alpha=1$, Eq.~(\ref{eq_free_tail_BA}) reproduces the known $k^{-2}$ for identical fermions.
The large-$|k|$ momentum tail for two fermionic anyons reads
\begin{eqnarray}
    \label{eq_free_tail_FA}
    \lim_{|k| \rightarrow \infty} n^{(\text{bd})}_{\alpha,-}(k) =
       \frac{4  {\cal{C}}^{(\text{bd})}_2}{k^2}  \cos^2 \left( \frac{\pi \alpha}{2} \right)
       - \frac{4  {\cal{C}}^{(\text{bd})}_2}{a_{\mathrm{sc}}k^3}
         \sin \left( \pi \alpha\right) 
        \nonumber \\
        + \frac{4  {\cal{C}}^{(\text{bd})}_2}{a^2_{\mathrm{sc}}k^4} \left[\sin^2 \left( \frac{\pi \alpha}{2} \right) - 2\cos^2 \left( \frac{\pi \alpha}{2} \right) \right] + {\cal{O}}(k^{-5}). \nonumber \\
\end{eqnarray}
For $\alpha=0$, Eq.~(\ref{eq_free_tail_FA}) reduces to the known $k^{-4}$ tail for identical bosons. For $\alpha = 1$, the expression only contains a $k^{-2}$ tail, as expected for two identical fermions~\cite{grosse2004exact,girardeau2003fermi, girardeau2004spinorFB, cui2016universal}.

It can be checked readily that Eqs.~(\ref{eq_free_tail_BA}) and (\ref{eq_free_tail_FA})
agree with the universal expressions reported in Tables~\ref{tab_momentum_tail} and \ref{tab_momentum_tail_fermion}, respectively. 
Applying the techniques developed in Ref.~\cite{paper1}, it can be shown that the first term in the square brackets in the second line of Eqs.~(\ref{eq_free_tail_BA}) and (\ref{eq_free_tail_FA}) is universal, i.e., fully governed by the boundary conditions given in Eqs.~(\ref{eq_BC_bosonicanyons}) and (\ref{eq_BC_fermionicanyons}). The second term in the square brackets in the second line of Eqs.~(\ref{eq_free_tail_BA}) and (\ref{eq_free_tail_FA}), in contrast, is non-universal for anyons with $0<\alpha<1$ and finite  $a_{\text{sc}}$
($|a_{\text{sc}}|$ not equal to zero or $\infty$). This non-universal contribution is discussed in more detail in the next section, where it is demonstrated that this contribution is modified in the presence of an external harmonic trapping potential. 

\subsection{Two identical  anyons under external harmonic confinement}
\label{sec_ho_anyons}

This section determines the tail of the momentum distribution for two identical bosonic and two identical fermionic anyons under external harmonic confinement with angular trapping frequency $\omega$. We start by determining the eigenenergies and eigenstates.
As in the case of the free-space system considered earlier, 
the center-of-mass motion can be separated off
 since the center-of-mass coordinate $Z$ is unchanged under the exchange of the coordinates of the two particles. We express energies and lengths in units of $\hbar \omega$ and $a_{\text{HO}}$, respectively, where $a_{\text{HO}}=\sqrt{\hbar/ (m\omega)}$. 
 Correspondingly, the eigenstates $\Psi^{(\text{HO})}_{M,\epsilon;\alpha,\pm}(z,Z)$ and eigenenergies $E_{M,\epsilon;\pm}$ are written as
 \begin{eqnarray}
 \Psi^{(\text{HO})}_{M,\epsilon;\alpha,\pm}(z,Z) = \Phi^{(\text{HO})}_{M}(Z)\psi^{(\text{HO})}_{\epsilon;\alpha,\pm}(z)
 \end{eqnarray}
 and
 \begin{eqnarray}
     \label{eq_ho_total}
     E_{M,\epsilon;\pm}=\left(M+\frac{1}{2}+\epsilon \right)\hbar \omega,
 \end{eqnarray}
 respectively. Here, $\epsilon \hbar \omega$ is the  energy associated with the relative wavefunction $\psi^{(\text{HO})}_{\epsilon;\alpha,\pm}(z)$.
 The center-of-mass wavefunction $\Phi^{(\text{HO})}_{M}(Z)$ can be written in terms of the Hermite polynomials $H_M(\sqrt{2}Z/a_{\text{HO}})$~\cite{mistakidis2023fewbody},
\begin{eqnarray}
    \label{eq_com_ho}
    \Phi^{(\text{HO})}_{M}(Z) = 
    \sqrt{\frac{\sqrt{2}}{2^M  M!a_{\text{HO}}\sqrt{\pi}}} \times \nonumber \\
    \exp\left(-\frac{Z^2}{a^2_{\text{HO}}}\right)H_M\left(\frac{\sqrt{2}Z}{a_{\text{HO}}}\right),
\end{eqnarray}
where $M$ denotes an integer quantum number, $M=0,1,\cdots$.
The center-of-mass energy is $(M+1/2)\hbar \omega$.

The anyonic eigenstates $\psi^{(\text{HO})}_{\epsilon;\alpha,\pm}(z)$ of the relative Hamiltonian can be obtained from the relative eigenstates $\psi^{(\text{HO})}_{\nu_{+}}(z)$ for two identical bosons and the relative eigenstates $\psi^{(\text{HO})}_{\nu_{-}}(z)$  for two identical fermions,  where the subscripts $\nu_+$ and $\nu_-$ denote non-integer quantum numbers that determine the relative  energies: $(2 \nu_++1/2)\hbar \omega$ for bosons
and $(2 \nu_-+3/2)\hbar \omega$ for fermions (see Appendix~\ref{sec_BF_HO}).
If $\nu_{\pm}$ are, for a given $\epsilon$ [see Eq.~(\ref{eq_ho_total}) for the definition of $\epsilon$],  chosen according to 
\begin{eqnarray}
\label{eq_nuplus_cond}
    \nu_+=\frac{\epsilon}{2}-\frac{1}{4}
\end{eqnarray}
and
\begin{eqnarray}
\label{eq_numinus_cond}
    \nu_-=\frac{\epsilon}{2}-\frac{3}{4},
\end{eqnarray}
it can be readily verified that the wave functions
\begin{eqnarray}
\label{eq_ho_anyon}
\psi^{(\text{HO})}_{\epsilon;\alpha, \pm}(z)=
{\cal{N}}(\alpha)
\bigg[
\cos \left( \frac{\pi \alpha}{2} \right)
\psi^{(\text{HO})}_{\nu_{\pm}}(z)+ \nonumber \\
\iu \sin \left(\frac{\pi \alpha}{2} \right)
\psi^{(\text{HO})}_{\nu_{\mp}}(z)
\bigg]
\end{eqnarray}
obey anyonic exchange statistics
and are eigenstates of the relative Hamiltonian.

\begin{figure}[h]
    \centering
    \includegraphics[width=0.8\linewidth]{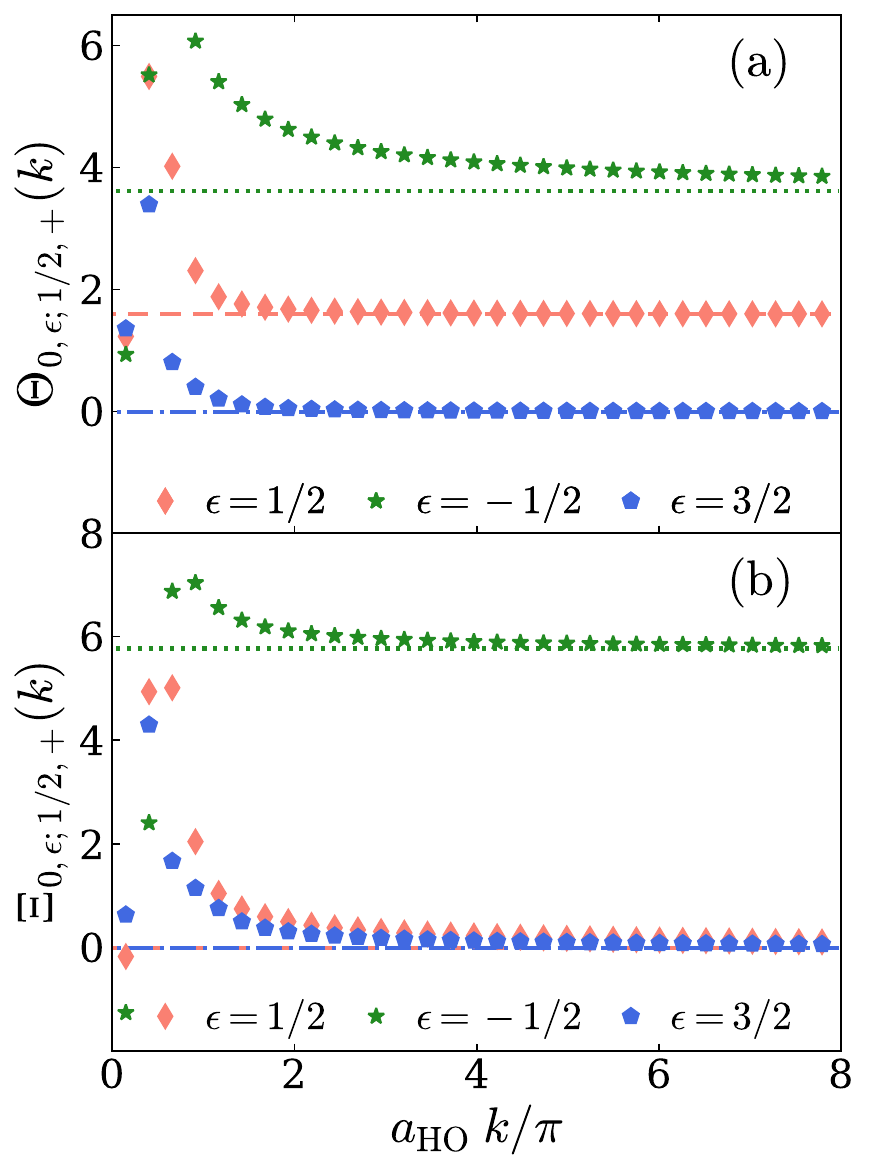}
    \caption{Analysis of the tail of the momentum distribution $n_{0,\epsilon;\alpha, +}^{(\text{HO})}(k)$   for two bosonic anyons with $\alpha = 1/2$ and $M=0$ (the center-of-mass coordinate is in the ground state) under external harmonic confinement for three different $\epsilon$. 
     The numerical results for the relative energies characterized by $\epsilon=+1/2$, $-1/2$, and $3/2$ are shown using diamonds, stars, and pentagons, respectively.
    For comparison, the analytical results are shown by dashed, dotted, and dash-dotted lines, respectively. 
    (a) Scaled momentum distribution $\Theta_{0,\epsilon;1/2,+}(k)$ [see Eq.~(\ref{eq_tail_k2_scaling})]. 
    (b) Shifted and scaled momentum distribution $\Xi_{0,\epsilon;1/2,+}(k)$ [see Eq.~(\ref{eq_tail_k3_scaling})]. 
  At large $k$, excellent agreement is found between the full numerical (scaled and shifted) momentum distribution and the analytically derived tail. 
    }
    \label{fig_momdist_any}
\end{figure}

\begin{figure}[h]
    \centering
    \includegraphics[width=1\linewidth]{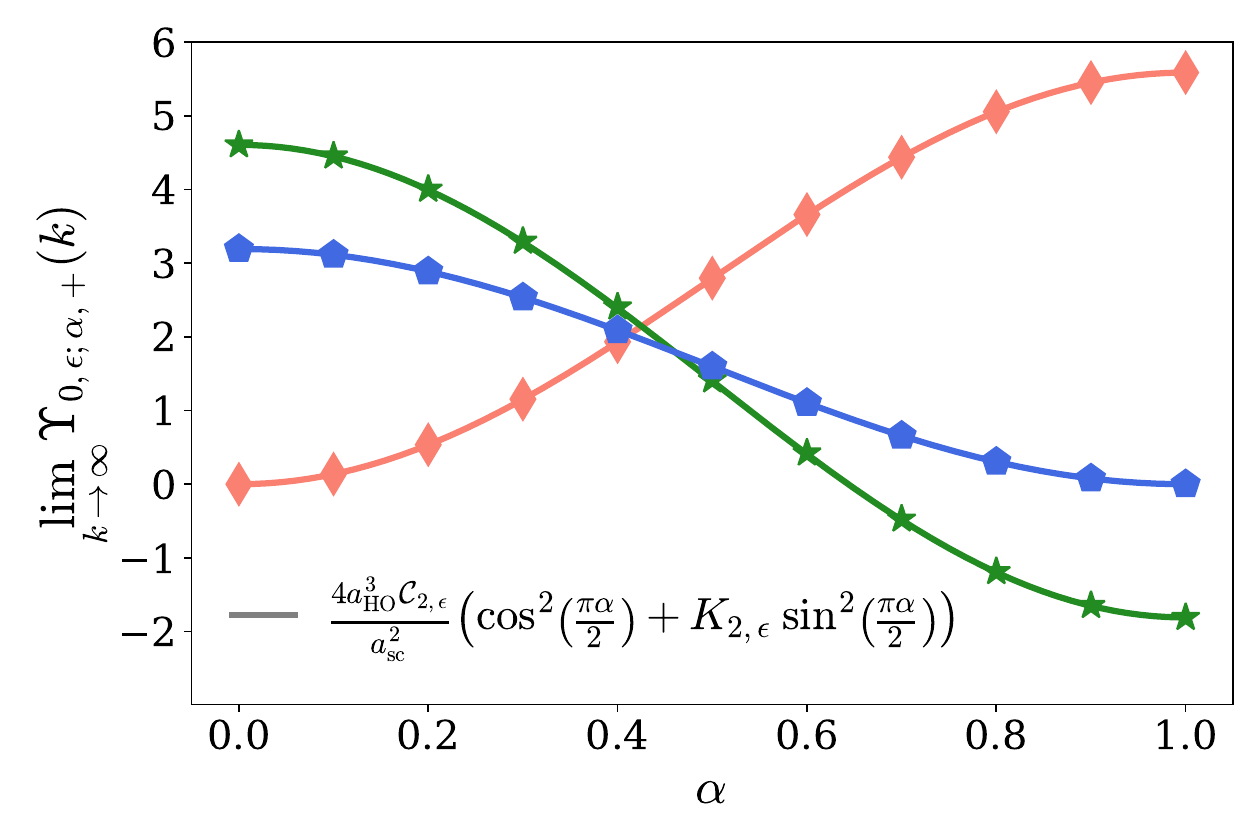}
    \caption{Analysis of the quantity $\lim_{k \rightarrow \infty}\Upsilon_{0,\epsilon;\alpha, +}^{(\text{HO})}(k)$ for two harmonically confined bosonic anyons as a function of $\alpha$. The numerical results for  $\epsilon=+1/2$, $-1/2$, and $3/2$, which are obtained for $a_\mathrm{HO}k = 100\pi$,  are shown using diamonds, stars, and pentagons, respectively. The solid lines show the corresponding analytical result (the expression can be found in the legend).
    The agreement between the numerical and analytical results is excellent.}  
    \label{fig_nkk4_alpha}
\end{figure}

The remainder of this section considers the $M=0$ case. 
To determine the momentum tail of two interacting identical anyons under harmonic confinement, we follow two routes: (i) We calculate $n_{0,\epsilon;\alpha,\pm}^{(\text{HO})}(k)$ numerically, using a  non-uniform spatial grid. Convergence is analyzed carefully by changing both the distribution of the grid points and the number of grid points. 
Symbols in Figs.~\ref{fig_momdist_any} and \ref{fig_nkk4_alpha} show our numerical results.
(ii) We calculate the tail of the momentum distribution analytically (see below for details).
Lines in Figs.~\ref{fig_momdist_any} and \ref{fig_nkk4_alpha} show our analytical results. The agreement between our numerical and analytical results at large $k$ is excellent. 
The analytical determination of the tail of the momentum distribution provides, as we now show, critical insights into   momentum tail contributions that arise from the interplay between the bosonic and fermionic parts of the relative anyonic wave function as well as the interplay between the relative and center-of-mass parts of the total anyonic wave function. Our derivation starts 
by Taylor-expanding the wavefunction $\Psi^{(\mathrm{HO})}_{0,\epsilon;\alpha,\pm}(z,Z)$ around $\xi=0$ up to order $\mathcal{O}(\xi^3)$,
where  $z_{1} = z_2 + \xi$.
Using $\xi$ and $z_2$ as our independent variables, we find for bosonic anyons (fermionic anyons can be treated analogously) 
\begin{eqnarray}
    \label{eq_expansionany_a_HO}
    \Psi^{(\mathrm{HO})}_{0,\epsilon;\alpha,+}(\xi,z_2+\xi/2) \underset{z_1\rightarrow z_2}{\longrightarrow} 
    \frac{\mathcal{M}(\frac{\epsilon}{2}-\frac{1}{4})}{a_\mathrm{HO}}\frac{(2\pi)^{1/4}e^{-\frac{z_2^2}{a_\mathrm{HO}^2}}}{\Gamma\left(\frac{3}{4} - \frac{\epsilon}{2}\right)} \nonumber \times \\
    \hat{S}^\dagger_{\alpha/2}(\xi)\bigg(1 - \frac{1}{4}\frac{\xi^2}{ a_\mathrm{HO}^2} - \frac{\xi z_2}{a_\mathrm{HO}^2} + \frac{1}{2}\frac{\xi^2 z_2^2}{a_\mathrm{HO}^4} \bigg) \times \nonumber \\
    \bigg[1 - \frac{\sqrt{2}}{J(\epsilon)}\frac{|\xi|}{a_\mathrm{HO}} - \frac{\epsilon}{2}\frac{\xi^2}{a_\mathrm{HO}^2}\bigg],\quad \quad
\end{eqnarray}
where $J(\epsilon)$ is equal to $ \frac{\Gamma(1/4-\epsilon/2)}{\Gamma(3/4-\epsilon/2)}$ and ${\cal{M}}(x)$ is defined in Appendix~\ref{sec_BF_HO}. 
The terms in the round brackets in the second line of  Eq.~(\ref{eq_expansionany_a_HO}) come from Taylor expanding  the center of mass wavefunction while the terms in the square brackets in the third line come from Taylor expanding  the relative wavefunction. 
Multiplying out the expressions in the second and third lines, the terms up to order $\mathcal{O}(\xi)$ read $1 - \frac{\sqrt{2}}{J(\epsilon)}\frac{|\xi|}{a_\mathrm{HO}}$. 
Using that $J(\epsilon)$ can be rewritten as $\sqrt{2} a_{\text{sc}}/a_{\text{HO}}$, we recognize that
these terms are equivalent to the logarithmic boundary condition given in Eq.~\eqref{eq_BC_bosons}.
This equivalence shows that these terms
give rise  to universal contributions to the tail of the momentum distributions~\cite{paper1}.
The higher-order terms are proportional to $\xi^{j-2}$  and $ \xi^{p}z_2^{q}$ ($j \ge4$ and $p+q\geq 2$).
These terms are not governed by the two-body zero-range boundary condition and give therefore rise to non-universal contributions to the momentum tail.
Inserting the expanded form of the total wave function into the definition of the momentum distribution,
we obtain the large-$|k|$ expression of the momentum distributions $n_{0,\epsilon;\alpha,\pm}^{(\text{HO})}(k)$ up to order ${\cal{O}}(k^{-4})$. The results are summarized in Tables~\ref{tab_momentum_tail_2bd_BA} and~\ref{tab_momentum_tail_2bd_FA} for bosonic and fermionic anyons, respectively.
The tables explicitly consider the case
where $a_{\text{sc}}$ is finite and $0<\alpha<1$ as well as the limiting cases of $|a_{\text{sc}}|=0$ and $|a_{\text{sc}}|=\infty$ (any $\alpha$),  $\alpha=0$ and $\alpha=1$ (any $a_{\text{sc}}$), 
and $\alpha=0$ or $1$ and $|a_{\text{sc}}|=0$ or $\infty$. 
The expressions in the tables are written in terms of ${\cal{C}}_{2,\epsilon}$ and $K_{2,\epsilon}$.
The two-body Tan contact $\mathcal{C}_{2,\epsilon}$  depends on the relative energy of the harmonically trapped two-anyon system (to emphasize the energy dependence, we added a subscript $\epsilon$ to the contact),
\begin{eqnarray}
    \mathcal{C}_{2,\epsilon} = \frac{2 \pi }{a_{\text{HO}}} \left[\frac{\mathcal{M}(\frac{\epsilon}{2} - \frac{1}{4})}{\Gamma(\frac{3}{4}-\frac{\epsilon}{2})} \right]^2.
\end{eqnarray} 
The dimensionless coefficient $K_{2,\epsilon}$  depends on both the relative and center-of-mass energy of the harmonically trapped two-anyon system, 
\begin{eqnarray}
     K_{2,\epsilon} =  \left(2\epsilon+\frac{3}{4}\right) \left(\frac{a_{\text{sc}}}{a_{\text{HO}}}\right)^2,
\end{eqnarray}
and is, just like the Tan contact,  independent of $\alpha$. For notational simplicity, we are suppressing the ``0''  center-of-mass ground state label of both $C_{2,\epsilon}$ and $K_{2,\epsilon}$.

In general, the tails of the momentum distribution contain terms proportional to $k^{-2}$, $k^{-3}$, and $k^{-4}$. To analyze these contributions term by term, we introduce the quantities $\Theta_{0,\epsilon;\alpha,+}(k)$, $\Xi_{0,\epsilon;\alpha,+}(k)$, and $\Upsilon_{0,\epsilon;\alpha,+}(k)$, which are defined so as to ``isolate'' the prefactors of the $k^{-2}$, $k^{-3}$, and $k^{-4}$ terms, respectively,
\begin{eqnarray}
    \label{eq_tail_k2_scaling}
    \Theta_{0,\epsilon;\alpha,+}(k) = \left[n^{(\mathrm{HO})}_{0,\epsilon;\alpha,+}(k)\right]a_\mathrm{HO}k^2,
\end{eqnarray}
\begin{eqnarray}
\label{eq_tail_k3_scaling}
    \Xi_{0,\epsilon;\alpha,+}(k) = \nonumber \\
    \left[n^{(\mathrm{HO})}_{0,\epsilon;\alpha,+}(k)-\frac{4\mathcal{C}_{2,\epsilon}}{k^2}\sin^2\left(\frac{\pi \alpha}{2}\right)\right]a^2_\mathrm{HO}k^3,
\end{eqnarray}
and
\begin{eqnarray}
    \label{eq_tail_k4_scaling}
    \Upsilon_{0,\epsilon;\alpha,+}(k) = \bigg[n^{(\mathrm{HO})}_{0,\epsilon;\alpha,+}(k)-\frac{4\mathcal{C}_{2,\epsilon}}{k^2}\sin^2\left(\frac{\pi \alpha}{2}\right) \nonumber \\
    -\frac{4\mathcal{C}_{2,\epsilon}}{a_\mathrm{sc}k^3}\sin\left(\pi \alpha\right)\bigg]a^3_\mathrm{HO}k^4.
\end{eqnarray}
The large-$|k|$ behavior of fermionic anyons can be analyzed analogously.
Figures~\ref{fig_momdist_any}(a) and \ref{fig_momdist_any}(b) show
$\Theta_{0,\epsilon;\alpha,+}(k)$ and $\Xi_{0,\epsilon;\alpha,+}(k)$, respectively, as a function of $k$ for $\alpha=1/2$ and three different relative energies, namely $\epsilon=-1/2$ (negative $a_{\text{sc}}$ corresponding to the ``molecular branch'' of bosonic anyons), 
$\epsilon=1/2$ (non-interacting bosonic anyons), and $\epsilon=3/2$ (infinitely repulsively interacting bosonic anyons).
The agreement between the numerical results (symbols) and the analytical results (lines)  is excellent. Specifically, the numerical results approach the analytical results at large $k$, thereby confirming the analytically derived $k^{-2}$ and $k^{-3}$ expressions of the tail. We note that the $k^{-3}$ contribution to the tail, which arises due to an interference between two wave function terms,  vanishes for $\epsilon =1/2$ and  $3/2$ because the quantity $\mathcal{C}_{2,\epsilon}/a_\mathrm{sc}$ goes to zero in these limits (see Appendixes~\ref{app_limitingcase_1/2} and~\ref{app_limitingcase_3/2} for details).

Figure~\ref{fig_nkk4_alpha} compares the $k^{-4}$ contribution to the momentum distribution tail  for two bosonic anyons under harmonic confinement   ($\epsilon = 1/2$, $-1/2$, and $3/2$). The  diamonds, stars, and pentagons  show the numerically calculated quantity $\lim_{k \rightarrow \infty}\Upsilon_{0,\epsilon;\alpha,+}(k)$ as a function of $\alpha$ for $\epsilon = 1/2$, $-1/2$, and $3/2$, respectively. The numerical results are obtained by evaluating the momentum distribution at $a_{\text{ho}}k=100 \pi$. For comparion, the  solid lines show our analytical result (see Table~\ref{tab_momentum_tail_2bd_BA} as well as Appendixes~\ref{app_limitingcase_1/2} and \ref{app_limitingcase_3/2}). The agreement is excellent. The analysis presented in Fig.~\ref{fig_nkk4_alpha} underlines that the $k^{-4}$ contribution to the  tail of the momentum distribution contains a non-universal contribution (if the behavior was universal, the three solid lines would coincide). 

In general (finite $a_{\text{sc}}$ and $0<\alpha<1$), the $k^{-4}$ term for bosonic anyons has two contributions: one that is proportional to $\cos^2(\pi \alpha/2)$ and one that is proportional to $\sin^2(\pi \alpha/2)$. Comparison of Table~\ref{tab_momentum_tail_2bd_BA} and Eq.~(\ref{eq_free_tail_BA}) shows that the former term is the same for bosonic anyons under harmonic confinement and for bosonic anyons in free space, while the latter term differs for bosonic anyons under harmonic confinement (with coefficient $K_{2,\epsilon}$) and for bosonic anyons in free space (with coefficient $-2$). 

 \begin{widetext}

 \begin{table}[h]
        \begin{tabular}{c|c|c|c}
             \enspace   & $\alpha = 0$  & $0<\alpha<1$  & $\alpha = 1$\\ \hline
        $|a_\mathrm{sc}| = 0$ & $\lim_{|a_{\text{sc}}|=0}\frac{4\mathcal{C}_{2,\epsilon}}{a_{\text{sc}}^2 k^4}$\,\footnote{In the $|a_{\text{sc}}|\rightarrow 0$ limit, ${\cal{C}}_{2,\epsilon}$ is directly proportional to $a_{\text{sc}}^2$ (see Appendix~\ref{app_limitingcase_3/2}).}
        & $\lim_{|a_{\text{sc}}|=0}\frac{4\mathcal{C}_{2, \epsilon}}{a_{\text{sc}}^2 k^4}\cos^2\left(\frac{\pi \alpha}{2}\right)\,^{\color{red}\mathrm{a}}$ & no algebraic tail \\
        $0<|a_\mathrm{sc}|<\infty$ & $\frac{4\mathcal{C}_{2,\epsilon}}{a^2_\mathrm{sc} k^4}$ & $\frac{4\mathcal{C}_{2,\epsilon}}{k^2}\sin^2\left(\frac{\pi \alpha}{2}\right)
    \!+\! \frac{4\mathcal{C}_{2,\epsilon}}{a_\mathrm{sc}k^3}\sin\left(\pi \alpha \right) \!+\! \frac{4\mathcal{C}_{2,\epsilon}}{a^2_\mathrm{sc} k^4}\left[\cos^2\left(\frac{\pi\alpha}{2}\right)\!+\!K_{2,\epsilon} \sin^2\left(\frac{\pi\alpha}{2}\right)\right]$ & $\frac{4\mathcal{C}_{2,\epsilon}}{k^2}
    \!+\! \frac{4\mathcal{C}_{2,\epsilon}}{a^2_\mathrm{sc} k^4}K_{2,\epsilon}$ \\
        $|a_\mathrm{sc}|^{-1} = 0$ & no algebraic tail & $\frac{4\mathcal{C}_{2,1/2 + 2n}}{k^{2}}\sin^2\left(\frac{\pi \alpha}{2}\right) \!+\! \frac{(7+16n)\mathcal{C}_{2,1/2 + 2n}}{a^2_\mathrm{HO}k^{4}}\sin^2\left(\frac{\pi \alpha}{2}\right)$ & $\frac{4\mathcal{C}_{2,1/2 + 2n}}{k^{2}} \!+\! \frac{(7+16n)\mathcal{C}_{2,1/2 + 2n}}{a^2_\mathrm{HO}k^{4}}$\\
        
        \end{tabular}
        \caption{Summary of the behavior of $\lim_{|k|\rightarrow\infty} n^{(\mathrm{HO})}_{0,\epsilon;\alpha, +}(k)$ for two identical bosonic anyons up to $\mathcal{O}(k^{-4})$. Terms that are proportional to $k^{-2}$ and $k^{-3}$ are universal. The sub-sub-leading terms (terms proportional to $k^{-4}$) contain universal and non-universal contributions: $k^{-4}$ terms that contain $K_{2,\epsilon}$ or ${\cal{C}}_{2,1/2+2n}$ are non-universal.
        The limiting statistics of $\alpha=0$ (identical bosons) and $\alpha=1$ (identical fermions) are reported in the second and fourth columns (the results for $\alpha=0$ can be found in Ref.~\cite{olshanii2003short}). Since the wave functions of two non-interacting bosons ($|a_{\text{sc}}|^{-1}=0$) and two non-interacting fermions ($|a_{\text{sc}}|=0$) do not possess any discontinuities, their momentum distributions do not possess algebraic tails.} \label{tab_momentum_tail_2bd_BA}
\end{table} 

\begin{table}[h]
        \begin{tabular}{c|c|c|c}
             \enspace   & $\alpha = 0$  & $0<\alpha<1$  & $\alpha = 1$\\ \hline
        $|a_\mathrm{sc}| = 0$ & no algebraic tail & $\lim_{|a_{\text{sc}}|=0}\frac{4\mathcal{C}_{2,\epsilon}}{a_{\text{sc}}^2 k^4}\sin^2\left(\frac{\pi \alpha}{2}\right)$\,\footnote{In the $|a_{\text{sc}}|\rightarrow 0$ limit, ${\cal{C}}_{2,\epsilon}$ is directly proportional to $a_{\text{sc}}^2$.} & $\lim_{|a_{\text{sc}}|=0}\frac{4\mathcal{C}_{2,\epsilon}}{a_{\text{sc}}^2 k^4}\,^{\color{red}\mathrm{a}}$ \\
        $0<|a_\mathrm{sc}|<\infty$ & $\frac{4\mathcal{C}_{2,\epsilon}}{k^2}
    \!+\! \frac{4\mathcal{C}_{2,\epsilon}}{a^2_\mathrm{sc} k^4}K_{2,\epsilon}$ & $\frac{4\mathcal{C}_{2,\epsilon}}{k^2}\cos^2\left(\frac{\pi \alpha}{2}\right)
    \!-\! \frac{4\mathcal{C}_{2,\epsilon}}{a_\mathrm{sc}k^3}\sin\left(\pi \alpha\right) \!+\! \frac{4\mathcal{C}_{2,\epsilon}}{a^2_\mathrm{sc} k^4}\left[\sin^2\left(\frac{\pi\alpha}{2}\right)\!+\!K_{2,\epsilon} \cos^2\left(\frac{\pi\alpha}{2}\right)\right]$ & $\frac{4\mathcal{C}_{2,\epsilon}}{a^2_\mathrm{sc} k^4}$ \\
        $|a_\mathrm{sc}|^{-1} = 0$ & $\frac{4\mathcal{C}_{2,1/2 + 2n}}{k^{2}} \!+\! \frac{(7+16n)\mathcal{C}_{2,1/2 + 2n}}{a^2_\mathrm{HO}k^{4}}$ & $\frac{4\mathcal{C}_{2,1/2 + 2n}}{k^{2}}\cos^2\left(\frac{\pi \alpha}{2}\right) \!+\! \frac{(7+16n)\mathcal{C}_{2,1/2 + 2n}}{a^2_\mathrm{HO}k^{4}}\cos^2\left(\frac{\pi \alpha}{2}\right)$ & no algebraic tail \\
        
        \end{tabular}
        \caption{Summary of the behavior of $\lim_{|k|\rightarrow\infty} n^{(\mathrm{HO})}_{0,\epsilon;\alpha, -}(k)$ for two identical fermionic anyons up to $\mathcal{O}(k^{-4})$. Terms that are proportional to $k^{-2}$ and $k^{-3}$ are universal. The sub-sub-leading terms (terms proportional to $k^{-4}$) contain universal and non-universal contributions: $k^{-4}$ terms that contain $K_{2,\epsilon}$ or ${\cal{C}}_{2,1/2+2n}$ are non-universal.
        The limiting statistics of $\alpha=0$ (identical fermions) and $\alpha=1$ (identical bosons) are reported in the second and fourth columns (the results for $\alpha=1$ can be found in Ref.~\cite{olshanii2003short}). Since the wave functions of two non-interacting bosons ($|a_{\text{sc}}|^{-1}=0$) and two non-interacting fermions ($|a_{\text{sc}}|=0$) do not possess any discontinuities, their momentum distributions do not possess algebraic tails.}
        \label{tab_momentum_tail_2bd_FA}
\end{table}
\end{widetext}

\section{Conclusion}
\label{sec_conclusion}
This paper considers two identical 1D anyons with zero-range interactions in the continuum, both in free space and under external harmonic confinement. We investigated the impact that the gauge phase, which defines the anyonic exchange symmetry, has on the system properties for both bosonic and fermionic anyons. Bound and scattering solutions were considered.
The free-space scattering phase shift was defined for bosonic and fermionic anyons by introducing regular and irregular reference functions that obey anyonic exchange statistics. Analytic continuation of the scattering solutions yields the anyon eigenstates with negative energy: there exists exactly one two-body bound state for bosonic anyons in free space and one two-body bound state for fermionic anyons in free space. We also determined compact expressions for the eigenenergies and eigenstates of two identical bosonic and two identical fermionic anyons under external harmonic confinement.

Additionally, we determined off-diagonal elements of observables that explicitly depend on both coordinates of the two-particle system, which are sensitive to the exchange statistics, i.e., the statistical parameter $\alpha$,  for the free-space and the harmonically-trapped two-anyon systems and 
 compared the off-diagonal correlations of bosonic and fermionic anyons. 
 The tails of the momentum distributions for bosonic and fermionic anyons were
 found to agree with the results of Ref.~\cite{paper1}. The existence of a non-universal contribution to the  $k^{-4}$ tail of the momentum distribution was  demonstrated explicitly for anyons in free space and under external harmonic confinement. 

 The frameworks and results presented in this paper might serve as a basis for investigating the dynamics of two identical anyons, which should contain fingerprints of the broken chiral symmetry.
 The two-body system considered in this work also serves as a building block for understanding the properties of systems consisting of more than two identical anyons.
It would, e.g., be interesting to develop a scattering theory for three identical anyons or a mixture of two identical anyons and a third distinguishable particle.
 
{\em{Acknowledgement:}}
We thank Jacob D. Norris for insightful discussions. R.H. and T.B. acknowledge support from Okinawa Institute of Science and Technology Graduate University and the Scientific Computing and Data Analysis
(SCDA) section of the Research Support Division at OIST. D.B. acknowledges support by the National Science Foundation through grant number PHY-2409311.

\bibliography{manuscript}

\appendix

\section{Derivation of Eqs.~(\ref{eq_BA_FA_mirrorWF_1}), (\ref{eq_BA_FA_mirrorWF_2}), and (\ref{eq_momentum_symm3})}
\label{app_derivation_some_symm}
Throughout this appendix, we assume that $\Psi_+(z_1,z_2)$ and
$\Psi_-(z_1,z_2)$ are real.
 Using Eq.~(\ref{eq_trick_sign}), we find 
\begin{eqnarray}
    \label{eq_chiralsymm_BA_FA}
    \Psi_{1-\alpha,-}(z_1,z_2) = \hat{S}^\dagger_{(1 - \alpha)/2}(z_1-z_2) \sign(z_1-z_2) \times \nonumber \\
    \Psi_{+}(z_1,z_2) = \nonumber \\
    \hat{S}^\dagger_{1/2}(z_1-z_2) \hat{S}^\dagger_{-\alpha/2}(z_1-z_2) \left(\iu \hat{S}_{1/2}(z_1-z_2)\right) \times \nonumber \\
    \Psi_{+}(z_1,z_2)= \nonumber \\
    \iu \hat{S}_{\alpha/2}(z_1-z_2) \Psi_+(z_1,z_2) = \nonumber \\
    \iu \left[\Psi_{\alpha, +}(z_1,z_2)\right]^*. \quad
\end{eqnarray}
We thus  obtain
\begin{eqnarray}
   \label{eq_appendixB} \Psi_{\alpha,+} (z_1,z_2) = \iu \left(\Psi_{1-\alpha,-}(z_1,z_2) \right)^*.
\end{eqnarray}
To proof Eq.~(\ref{eq_momentum_symm3}), we use Eq.~(\ref{eq_appendixB}) in  $n_{\alpha, +}(k)$:
\begin{eqnarray}
    \label{eq_proof_momdist_chiralmirror_1}
    n_{\alpha, +}(k) = \int^{\infty}_{-\infty} dz_2  \times \nonumber \\
    \bigg{|} \int^{\infty}_{-\infty} dz_1  e^{-\iu k (z_1 - z_2)}\Psi_{\alpha,+} (z_1,z_2) \bigg{|}^2 = \nonumber \\
    \int^{\infty}_{-\infty} dz_2  \times \nonumber \\
    \bigg{|} \int^{\infty}_{-\infty} dz_1  e^{-\iu k (z_1 - z_2)}\iu\left(\Psi_{1-\alpha,-}(z_1,z_2) \right)^* \bigg{|}^2 = \nonumber \\
    \int^{\infty}_{-\infty} dz_2  \times \nonumber \\
    \bigg{|} \int^{\infty}_{-\infty} dz_1  e^{-\iu k (z_1 - z_2)}\hat{S}_{1-\alpha}(z_1-z_2) \Psi_{-}(z_1,z_2) \bigg{|}^2 = \nonumber \\
    \int^{\infty}_{-\infty} dz_2  \times \nonumber \\
    \bigg{|} \int^{\infty}_{-\infty} dz_1  e^{\iu k (z_1 - z_2)}\hat{S}^\dagger_{1-\alpha}(z_1-z_2) \Psi_{-}(z_1,z_2) \bigg{|}^2 = \nonumber \\
    n_{1-\alpha, -}(-k). \nonumber \\
\end{eqnarray}
The expression for $n_{\alpha, -}(k)$ can be derived analogously.

\section{Relative eigenstates of two identical bosons and two identical fermions}
\label{sec_BF_HO}
This section reviews the eigenenergies and eigenstates of two identical bosons and two identical fermions under external harmonic confinement with angular trapping frequency $\omega$.
The even-parity function
$\psi^{(\text{HO})}_{\nu_{+}}(z)$
can be written in terms of the confluent hypergeometric  function $U$~\cite{busch1998two},
\begin{eqnarray}
\label{eq_relative_ho_bosonic}
\psi^{(\text{HO})}_{\nu_+}(z)= \frac{{\cal{M}}({\nu}_+)}{\sqrt{a_\mathrm{HO}}} \exp\left( -\frac{z^2}{4a^2_{\text{HO}}} \right) U \left(-\nu_+,\frac{1}{2}, \frac{z^2}{2a_{\text{HO}}^2} \right),\nonumber \\
\end{eqnarray}
where ${\cal{M}}({\nu}_+)$ denotes an energy-dependent normalization constant,
\begin{eqnarray}
    {\cal{M}}({\nu}_+) = \left(\int^{\infty}_{-\infty} |\psi^{(\text{HO})}_{\nu_+}(z)|^2 dz \right)^{-1/2}.
\end{eqnarray}
The allowed values of the non-integer quantum number $\nu_+$ depend on 
the even-parity coupling constant $g_+$ and are implicitly defined by  the transcendental equation
\begin{eqnarray}
-\frac{\mu g_+ a_\mathrm{HO}}{\hbar^2} 
= \frac{\sqrt{2}\Gamma\left(1/2 - \nu_+\right)}{\Gamma\left(-\nu_+\right)}.
\end{eqnarray}
The relative eigenstates $\psi^{(\text{HO})}_{\nu_-}(z)$
for two fermions are obtained via the Bose-Fermi mapping through the application of the $\sign(z)$ function,
\begin{eqnarray}
    \label{eq_relative_ho_fermionic}
    \psi^{(\text{HO})}_{\nu_-}(z)= \sign(z)\psi_{\nu_+}^{(\text{HO})}(z),
\end{eqnarray}
provided \begin{eqnarray}
    \nu_+ = \nu_- + \frac{1}{2}.
\end{eqnarray}
The non-integer quantum number $\nu_-$ is a solution to the transcendental equation
\begin{eqnarray}
\frac{\hbar^2}{\mu}\frac{a_\mathrm{HO}}{g_-} 
= \frac{\sqrt{2}\Gamma\left(-\nu_-\right)}{\Gamma\left(-1/2 - \nu_-\right)}.
\end{eqnarray}

\section{Bosonic anyons:  $|a_\mathrm{sc}| = 0$}
\label{app_limitingcase_3/2}
This appendix  considers the limiting case $|a_{\text{sc}}|=0$ (infinitely repulsive or infinitely attractive bosonic anyons under harmonic confinement), which corresponds to $\epsilon = 3/2 + 2n$ or $\nu_+ = 1/2 + n$ ($n=0,1,\cdots$). The normalization of the relative wavefunction is~\cite{daily2012occupation}
\begin{eqnarray}
    \mathcal{M}\left(\frac{1}{2} + n\right) = \frac{\left(\frac{2}{\pi}\right)^{1/4}}{\left[\left(\frac{3}{2}\right)_n(1)_n\right]^{1/2}},
\end{eqnarray}
where $(\lambda)_n$ is the Pochhammer function for  positive integer $n$. 
In this limit,  $a_{\text{sc}}$,  ${\cal{C}}_{2,\epsilon}$, and $K_{2,\epsilon}$  go to zero. We find
\begin{eqnarray}
    \lim_{|a_\mathrm{sc}|=0} \frac{{\cal{C}}_{2,\epsilon}}{a_{\text{sc}}} = 0,
\end{eqnarray}
which implies that the $k^{-3}$ tail is absent.
The ratio ${\cal{C}}_{2,\epsilon}/a_{\text{sc}}^2$, in contrast, is finite: 
\begin{eqnarray}
    \lim_{|a_\mathrm{sc}|=0} 
    \frac{\mathcal{C}_{2,\epsilon}}{a_\mathrm{sc}^2} = \lim_{\epsilon\rightarrow\frac{3}{2} + 2n} 
    \frac{2\mathcal{C}_{2,\epsilon}}{a_\mathrm{HO}^2 [J\left(\epsilon\right)]^2} =  \nonumber \\ 
    \frac{\sqrt
    \frac{2}{\pi}}{a_\mathrm{HO}^3}\frac{\left(\frac{3}{2}\right)_n}{(1)_n}.\quad \quad \quad \quad
\end{eqnarray}
Lastly, we find
\begin{eqnarray}
    \lim_{|a_\mathrm{sc}|=0} 
    \frac{\mathcal{C}_{2,\epsilon}K_{2,\epsilon}}{a_\mathrm{sc}^2} =  0.
\end{eqnarray}
Putting everything together, the  momentum tail, up to order $\mathcal{O}(k^{-4})$, is 
\begin{eqnarray}
    n_{0, \frac{3}{2} +2n;\alpha, +}(k) = \frac{4\left(\frac{3}{2}\right)_n}{(1)_n}\frac{\sqrt
    \frac{2}{\pi}}{a_\mathrm{HO}^3 k^{4}}\cos\left(\frac{\pi \alpha}{2}\right)^2.
\end{eqnarray}
The result is universal.

\section{Bosonic anyons:  $|a_\mathrm{sc}|^{-1} = 0$}
\label{app_limitingcase_1/2}
This appendix
 considers the limiting case $|a_\mathrm{sc}|^{-1} = 0$ (two non-interacting bosonic anyons under harmonic confinement), which corresponds to $\epsilon = 1/2 + 2n$ ($n=0,1,\cdots$).   The normalization of the wavefunction is~\cite{daily2012occupation}
\begin{eqnarray}
    \mathcal{M}\left(n\right) = \frac{(2\pi)^{-1/4}}{\left(\left(\frac{1}{2}\right)_n(1)_n\right)^{1/2}}.
\end{eqnarray}
Evaluating Eq.~\eqref{eq_twobody_contact}, the Tan contact becomes
\begin{eqnarray}
    \mathcal{C}_{2,\frac{1}{2} + 2n} = \frac{\sqrt{\frac{2}{\pi}}}{a_\mathrm{HO}} \frac{\left(\frac{1}{2}\right)_n}{\left(1\right)_n}.
\end{eqnarray}
Since  $1/a_{\text{sc}}=0$, we have 
\begin{eqnarray}
    \lim_{|a_\mathrm{sc}|^{-1}=0} \frac{{\cal{C}}_{2,\epsilon}}{a_{\text{sc}}} = 0,
\end{eqnarray}
which implies that the $k^{-3}$ tail  is absent. Likewise, we have
\begin{eqnarray}
    \lim_{|a_\mathrm{sc}|^{-1}=0} 
    \frac{\mathcal{C}_{2,\epsilon}}{a_\mathrm{sc}^2} = 0.
\end{eqnarray}
On the other hand,  $K_{2,\epsilon}$ diverges in this limit. The ratio $K_{2,\epsilon}/a_{\text{sc}}^2$, however, is well behaved: 
\begin{eqnarray}
    \lim_{|a_\mathrm{sc}|^{-1}=0} 
    \frac{K_{2,\epsilon}}{a_\mathrm{sc}^2} =  \frac{\frac{7}{4}+4n}{a_\mathrm{HO}^2}.
\end{eqnarray}
Putting everything together, the momentum tail, up to order $\mathcal{O}(k^{-4})$, is
\begin{eqnarray}
    n_{0, \frac{1}{2} +2n;\alpha, +}(k) = \frac{4C_{2,\frac{1}{2} + 2n}}{k^2}\sin^2\left(\frac{\pi \alpha}{2}\right) + \nonumber \\
    \frac{4\mathcal{C}_{2,\frac{1}{2} + 2n}(\frac{7}{4}+4n)}{a_\mathrm{HO}^2k^4}\sin^2\left(\frac{\pi \alpha}{2}\right).
\end{eqnarray}
The $k^{-2}$ contribution is universal while  the $k^{-4}$ contribution, which possesses an $n$-dependent coefficient, is non-universal.
\end{document}